\documentstyle[twocolumn,psfig]{article}

\def\epsfig{\psfig}

\makeatletter
\advance \topmargin by -\headheight
\advance \topmargin by -\headsep
\evensidemargin \oddsidemargin
\newlength{\figurewidth}
\newlength{\dblfigurewidth}
\def\thebibliography#1{\list
 {\arabic{enumi}.}{\settowidth\labelwidth{[#1]}\leftmargin\labelwidth
 \advance\leftmargin\labelsep
 \usecounter{enumi}}
 \parsep 0pt\itemsep \parsep
 \parskip 0pt
 \def\newblock{\hskip .11em plus .33em minus .07em}
 \sloppy\clubpenalty4000\widowpenalty4000
 \sfcode`\.=1000\relax\small}
\def\lesssim{\mathrel{\lower2.5pt\hbox{$\textstyle<$}\atop\raise2.5pt\hbox{$\textstyle\sim$}}}

\def\pacs#1{}

\def\onlinecite#1{\citen{#1}}
\long\def\@caption#1[#2]#3{\par\addcontentsline{\csname
  ext@#1\endcsname}{#1}{\protect\numberline{\csname
  the#1\endcsname}{\ignorespaces #2}}\begingroup
    \@parboxrestore
    \small
    \@makecaption{\csname fnum@#1\endcsname}{\ignorespaces #3}\par
  \endgroup}
\def\@communication{}
\def\@authoraddress{}  \def\@title{} \def\@date{} \def\@preprint{}
\def\and{\unskip, }
\def\preprint#1{%
\def\@preprint{\hbox{#1}}%
}
\def\communication#1{%
\def\@communication{\hbox{#1}}%
}
\def\title#1{\gdef\@title{{\LARGE\bf\ignorespaces#1\vskip.5\baselineskip}}}
\def\author#1{\expandafter\def\expandafter\@authoraddress\expandafter
{\@authoraddress %
{\dimen0=-\prevdepth \advance\dimen0 by23pt
\nointerlineskip \Large\sc
\vrule height\dimen0 width0pt\relax\ignorespaces#1\par
}%
}%
}
\def\address#1{\expandafter\def\expandafter\@authoraddress\expandafter
{\@authoraddress{\vskip.5\baselineskip\ignorespaces#1\par}}}
\def\date#1{\gdef\@date{{\rm(\ignorespaces#1\unskip)\par}}}
\def\maketitle{\par
\begingroup
\let\cite\@bylinecite
\def\thefootnote{\fnsymbol{footnote}}%
\if@twocolumn
\twocolumn[\@maketitle\vskip2pc]%
\else
\newpage
\global\@topnum\z@ %
\@maketitle
\fi
\thispagestyle{plain}\@thanks
\endgroup
\def\thefootnote{\arabic{footnote}}%
\setcounter{footnote}{0}%
\let\maketitle\relax \let\@maketitle\relax
\let\@thanks\relax \let\@authoraddress\relax \let\@title\relax
\let\@date\relax \let\thanks\relax
}
\def\@maketitle{%
\begin{flushleft}
\@preprint \vskip 2\baselineskip
\@title
\@authoraddress \vskip \baselineskip
\@date
\vskip -1\baselineskip
\end{flushleft}
}
\def\thesection       {\Roman{section}}
\def\p@section        {}
\def\thesubsection    {\Alph{subsection}}
\def\p@subsection     {\thesection\,}

\def\p@subsubsection  {\thesection\,\thesubsection\,}

\def\acknowledgments{\section*{ACKNOWLEDGMENTS}}

\newif\if@mainhead
\let\reset@font\relax
\def\section{\@mainheadtrue
\@startsection {section}{1}{\z@}{-0.5cm plus-1ex minus
 -1ex}{0.2cm plus1ex minus1ex}{\reset@font\bf\centering}}
\def\subsection{\@mainheadfalse
\@startsection{subsection}{2}{\z@}{-0.8cm plus-1ex minus
 -.2ex}{0.5cm plus1ex minus.2ex}{\reset@font\small\bf\centering}}
\def\subsubsection{\@mainheadfalse
\@startsection{subsubsection}{3}{\z@}{-.8cm plus-1ex minus
 -.2ex}{0.5cm plus1ex minus.2ex}{\reset@font\small\it\centering}}
\def\paragraph{\@mainheadfalse
\@startsection{paragraph}{4}{\parindent}{\z@}{-1em}{\reset@font
\normalsize\it}}
\def\subparagraph{\@mainheadfalse
\@startsection{subparagraph}{4}{\parindent}{3.25ex plus1ex minus
 .2ex}{-1em}{\reset@font\normalsize\bf}}
\def\ps@myheadings{\let\@mkboth\@gobbletwo
\def\@oddhead{\hbox{}\rightmark \hfill \rm\thepage}
\def\@oddfoot{}\def\@evenhead{\rm \thepage\hfil\sl\leftmark\hbox
{}}\def\@evenfoot{}\def\sectionmark##1{}\def\subsectionmark##1{}}
\@ifundefined{citeleft}{\let\citeleft=[}{}
\@ifundefined{citeright}{\let\citeright=]}{}
\@ifundefined{citemid}{\def\citemid{,\penalty\@medpenalty\ }}{}
\@ifundefined{citepunct}{
   \def\citepunct{,\penalty\@highpenalty\hskip.13emplus.1emminus.1em}%
  }{}
\@ifundefined{citeform}{\def\citeform{}}{}
\let\oc@verbo\relax
\@ifundefined{DeclareOption}{}%
{ \toks@={\def\oc@verbo#1#2#3#4{}}
  \DeclareOption{verbose}{\the\toks@}
  \DeclareOption{ref}{\def\citeleft{[Ref.\penalty\@M\ }}
  \DeclareOption{nospace}{\def\citepunct{,\penalty\@highpenalty}}
  \DeclareOption{space}{\def\citepunct{,\penalty\@highpenalty\ }}
  \ProcessOptions
  \ProvidesPackage{overcite}[1994/07/11 \space  v 3.3]
}
\edef\citen{\noexpand\protect \expandafter\noexpand\csname citen \endcsname}
\@namedef{citen }#1{%
\edef\@tempa{\@ignspaftercomma,#1, \@end, }
\edef\@tempa{\expandafter\@ignendcommas\@tempa\@end}%
\if@filesw \immediate \write \@auxout {\string \citation {\@tempa}}\fi
\@tempcntb\m@ne    
\let\@h@ld\relax   
\let\@citea\@empty 
\let\@celt\over    
\def\@cite@list{}
\@for \@citeb:=\@tempa \do{\@make@cite@list}
\@tempcnta\m@ne    
\let\@celt\@compress@cite \@cite@list 
\@h@ld}
\begingroup \catcode`\_=8 
\toks@={
\def\@make@cite@list{%
 \expandafter\let \expandafter\@B@citeB
          \csname b@\@citeb\@extra@b@citeb \endcsname
 \ifx\@B@citeB\relax 
    \@citea {\bf ?}\let\@citea\citepunct
    \@warning {Citation `\@citeb' on page \thepage\space undefined}%
    \oc@verbo \global\@namedef{b@\@citeb\@extra@b@citeb}{?}%
 \else 
    \ifcat _\ifnum\z@<0\@B@citeB _\else A\fi 
       \@tempcnta\@B@citeB \relax
       \ifnum \@tempcnta>\@tempcntb 
          \edef\@cite@list{\@cite@list \@celt{\@B@citeB}}%
          \@tempcntb\@tempcnta
       \else 
          \edef\@cite@list{\expandafter\@sort@celt \@cite@list \@gobble @}%
       \fi
    \else 
       \@citea \citeform{\@B@citeB}%
       \let\@citea\citepunct
 \fi\fi}
}
\expandafter \endgroup \the\toks@  
\def\@compress@cite#1{
  \advance\@tempcnta\@ne 
  \ifnum #1=\@tempcnta   
     \ifx\@h@ld\relax    
        \edef \@h@ld{\@citea \citeform{#1}}%
     \else               
        \edef \@h@ld{\hbox{--}\penalty\@highpenalty \citeform{#1}}%
     \fi 
  \else   
     \@h@ld \@citea \citeform{#1}\let\@h@ld\relax
  \fi \@tempcnta#1\let\@citea\citepunct
}
\def\@sort@celt#1#2{\ifx \@celt #1
     \ifnum #2<\@tempcnta 
        \@celt{#2}%
        \expandafter\expandafter\expandafter\@sort@celt 
     \else 
        \@celt{\number\@tempcnta}\@celt{#2}
  \fi\fi}
\edef\cite{\noexpand\protect\expandafter\noexpand\csname cite \endcsname}
\@namedef{cite }{\leavevmode\unskip\@ifnextchar[{\@tempswatrue\@citew}%
            {\@tempswafalse\@citex}}
\def\@citew[#1]#2{\ifnum\lastpenalty=\z@ \penalty\@highpenalty \fi
   \ \citeleft{\multiply\@highpenalty 3 
   \citen{#2}}\citemid#1\citeright\spacefactor\@m}
\def\@citex#1{\begingroup \leavevmode \@tempcnta\@m \unskip
  \/
  \def\@tempa{\@cite{\citen{#1}}\spacefactor\@tempcnta\endgroup}%
  \futurelet\@tempb\@citey}%
\def\@citey{\let\@tempc\@tempa
   \ifx\@tempb.\ifnum\spacefactor>2999 \let\@tempb\relax\fi\let\@tempc\@citez
   \else\ifx\@tempb,\let\@tempc\@citez
   \else\ifx\@tempb:\let\@tempc\@citez 
   \else\ifx\@tempb;\let\@tempc\@citez 
   \fi\fi\fi\fi
   \@tempc}%
\def\@citez#1{\@tempcnta\sfcode`#1\@tempb\futurelet\@tempb\@citey}%
\def\@cite#1{$\m@th^{\hbox{\@ove@rcfont#1}}$}
\@ifundefined{selectfont}{
   \def\@ove@rcfont{\the\scriptfont\z@ \def\bf{\the\scriptfont\bffam}}
  }{
  \@ifundefined{fontsize}{
    \def\@ove@rcfont{\the\scriptfont\z@}\let\bf\relax 
    }{
    \def\@ove@rcfont{\csname S@\f@size\endcsname
        \fontsize{\sf@size}{\baselineskip}\selectfont}
 }}
\def\@ignspaftercomma#1, {\ifx\@end#1\@empty\else
   #1,\expandafter\@ignspaftercomma\fi}
\def\@ignendcommas,#1,\@end{#1}
\@ifundefined{@extra@b@citeb}{\def\@extra@b@citeb{}}{}
\makeatother

\begin{document}

\preprint{chem-ph/9507003}

\title{Kinetic and thermodynamic analysis of proteinlike
  heteropolymers: Monte Carlo histogram technique}

\author{Nicholas D. Socci and Jos\'e Nelson Onuchic}

\address{Department of Physics, University of California at San Diego,
  La~Jolla, California 92093-0319}

\date{In press {\em J.~Chem.~Phys.}}

\maketitle

\markright{{\sc Socci \& Onuchic}\hfill {\em Kinetic and thermodynamic
    analysis of proteinlike heteropolymers}}

\begin{abstract}
  Using Monte Carlo dynamics and the Monte Carlo Histogram Method, the
  simple three-dimensional 27 monomer lattice copolymer is examined in
  depth. The thermodynamic properties of various sequences are
  examined contrasting the behavior of good and poor folding
  sequences. The good (fast folding) sequences have sharp well-defined
  thermodynamic transitions while the slow folding sequences have
  broad ones. We find two independent transitions: a collapse
  transition to compact states and a folding transition from compact
  states to the native state. The collapse transition is second
  order-like, while folding is first order. The system is also studied
  as a function of the energy parameters. In particular, as the
  average energetic drive toward compactness is reduced, the two
  transitions approach each other. At zero average drive, collapse and
  folding occur almost simultaneously; i.e., the chain collapses
  directly into the native state. At a specific value of this energy
  drive the folding temperature falls below the glass point,
  indicating that the chain is now trapped in local minimum. By
  varying one parameter in this simple model, we obtain a diverse
  array of behaviors which may be useful in understanding the
  different folding properties of various proteins.
\end{abstract}

\section{INTRODUCTION}

Simple models are one powerful theoretical tool for the study of
complex systems. The idea is to remove all but the essentials from the
original system which will ideally allow for a more complete analysis.
There is often a trade-off between the complexity of the model (or how
faithfully it represents the system of interest) and the thoroughness
of the analysis. In the case of protein folding, research has spanned
the entire spectrum from all-atom molecular dynamics with
solvent\cite{Levitt88,Tobias91,Daggett92,Gilson94} to complete
enumeration of simple lattice polymer systems,\cite{Chan93a,Chan94}
with many works in between these two extremes. Naturally the more
realistic simulations do not yield results as thorough as the simpler
ones. In the all-atoms simulations a large amount of supercomputer
time is required for runs of hundreds of picoseconds of a single
protein molecule (plus solvent). In contrast, high-end workstations
can be used to simulate lattice polymers. Many different sequences can
be simulated over a range of temperatures for time scales comparable
to the folding time. We do not want to imply that one set of
techniques is superior than the other, but rather in studying a system
as complex as proteins many different approaches are necessary.  In
fact the two limits complement each other. Simple systems permit
detailed analysis while the more complex systems allow for contact
with real proteins.  Connecting these two limits would allow for a
more through analysis of real protein systems. Such an analysis has
been recently carried out.\cite{Bryngelson95,Onuchic95}

In a previous work\cite{Socci94b} we examined the kinetics of a simple
three-dimensional lattice polymer system. This system is too large for
exact enumeration studies but is still small enough for detailed
analysis. Many studies of lattices models seem to focus either on
thermodynamics or on kinetics, considering each in isolation. However,
as previously shown\cite{Socci94b} and shown here, a combined approach
that considers both the kinetics and thermodynamics of the same system
is important in understanding the model in detail. We have determined
that there is an important relation between kinetics (the ``glass
transition'') and thermodynamics (the folding temperature) in
determining whether a sequence will fold or not, an idea that was
advanced by Bryngelson and Wolynes,\cite{Bryngelson87,Bryngelson89}
and later explored by Leopold and Onuchic.\cite{Leopold92} In this
work we continue the study of this system, examining both
thermodynamics and kinetics in greater detail.

Some of the earliest work on the thermodynamics of protein-like
lattice polymers has been performed by Chan, Lau and
Dill.\cite{Lau89,Chan91a} They examined short chains in two dimensions
for which it is possible to enumerate all conformations.  This allowed
them to calculate any thermodynamic quantity by simply summing over
states. By measuring a variety of parameters, such as the average
compactness, number of contacts and hydrophobic core, they found a
distinct difference between folding and non-folding sequences.
Although exact enumeration studies are extremely powerful, they are
limited to small chains, usually in two dimensions.  In three
dimensions many have studied the 27 monomer system on a simple cubic
lattice which has a maximally compact shape of a
$3\!\times\!3\!\times\!3$ cube.  It is not possible to enumerate all
conformations, but it is still possible to enumerate all {\em compact}
conformations (what is often referred to as the {\em cube spectrum})
and then calculate approximate thermodynamics using just the cube
states.\cite{Shakhnovich90b,Sali94} However, as we show in this work,
care must be taken in using this technique since the cube states are
not an accurate approximation to the full density of states; in
particular there are many low energy (i.e., thermodynamically
relevant) states that are not cubes.

For longer chains, one can use the standard Monte Carlo
technique\cite{Metropolis53} for calculating thermodynamic properties.
Skolnick, Sikorski and
Kolinski\cite{Sikorski90,Skolnick90a,Skolnick91} use a dynamic Monte
Carlo method to study folding of realistic protein-like structures on
a diamond lattice. The word dynamic here is used to indicated that the
move set has been selected in such a way that the actual time course
of folding is as realistic as possible. Note that this is not
necessary for calculating thermodynamic quantities, since it is only
necessary that the moves satisfy detailed balance. In fact, O'Toole
and Panagiotopoulos\cite{OToole92} found that using the dynamic moves
for the 3D cubic lattice system led to sampling problems. At low
temperature they observed a hysteresis for the average energy. This is
probably caused by trying to sample below the glass transition. This
transition was predicted\cite{Bryngelson87} and explicitly
shown\cite{Socci94b} to exist in these simple lattice systems.
O'Toole and Panagiotopoulos were able to circumvent this problem by
using a more sophisticated sampling procedure based on the Rosenbluth
and Rosenbluth chain growth algorithm.\cite{Rosenbluth55} They studied
chains as long as 48 monomers and have used this technique to study
thermodynamically significant low energy conformations.\cite{OToole93}
Others have used Monte Carlo techniques to examine the effects of
various potential functions on the kinetics and thermodynamics of
two-dimensional lattice polymer systems.\cite{Camacho93}

These previous studies used the standard Monte Carlo technique.  The
simulation is run at a given temperature and various averages are
computed. To obtain thermodynamic quantities for a different
temperature, another simulation (at the new temperature) is performed.
However, it is possible to extract information about temperatures
other than the simulation temperature using a technique often called
the {\em Monte Carlo Histogram
  Method}.\cite{Ferrenberg88,Ferrenberg89} Using this technique one
can calculate an approximate density of states for the system, which
can be used to calculate any thermodynamic quantity of interest over a
range of temperatures. Because of the small system sizes used, we can
obtain accurate results over a broad range of temperatures. In
particular, we can extrapolate into the glass region where running
normal simulations becomes extremely difficult and time consuming. The
technique also facilitates the finding of peaks or zeros of various
thermodynamic functions. It can be used to calculate extensive
quantities like the free energy or entropy of the system, which are
difficult to extract by the standard Monte Carlo procedure. One can
not only vary the temperature, but also the various parameters in the
potential. One can examine how the system behaves thermodynamically at
a range of parameter values without the need to run new simulations.
This in-depth analysis of the thermodynamics has enabled us, along
with several others, to begin to connect the behavior and properties
of these simple model systems with those of real
proteins.\cite{Bryngelson95,Onuchic95}

\section{MODEL AND METHODS}

\subsection{Definition of model and dynamics}

The model studied was a simple three-dimensional cubic lattice
polymer. All chains were 27 monomers long with two different monomer
types.  The potential was a contact interaction between monomers that
are nearest neighbors on the lattice (but that are not linked
covalently). The energy of a contact was $E_l$ for a pair of the same
monomer type and $E_u$ for a pair of different monomers. This model
was the same one used in our previous study\cite{Socci94b} (which can
be consulted for a more detailed explanation of the model).  The
energy
function is:
\begin{equation}
  E = N_l E_l + N_u E_u,
\label{eq:potential}
\end{equation}
where $N_l$ is the number of contacts between monomers of the same
type (like contacts) and $N_u$ the number of contacts between monomers
of different types.

Although it is easiest to express the energy function in terms of the
$E_l$ and $E_u$ variables, some properties of the model are more
clearly understood by considering an equivalent set of parameters,
$E_{\rm
  avg}$ and $\Delta$, defined as follows:
\begin{eqnarray}
  E_{\rm avg} &=& \frac{1}{2}\left(E_l+E_u\right)\nonumber\\ \Delta
  &=& \left(E_u - E_l\right).
  \label{eq:eavgdelta}
\end{eqnarray}
$E_{\rm avg}$ represents the overall drive toward forming contacts or
compacting the chain. If it is less than zero, contact formation will
be favored. $\Delta$\ determines the heterogeneity of the
heteropolymer. In the limit that $\Delta=0$ the model becomes a
homopolymer. In our previous work,\cite{Socci94b} $E_{\rm avg}=-2$ and
$\Delta=2$, giving values -3 and -1 for $E_l$ and $E_u$
respectively.\cite{Note08} This insures that the chain collapses
rapidly, compared to folding, and that the minimum energy state is a
maximally compact cube for the sequences considered here. When we say
the chain has folded we mean it is in the native state. This is
distinguished from collapse, which refers to chains that can be in any
compact conformation. The same parameters were used for part of the
results presented here. In addition we show how the model behaves as
these parameters are varied.

\begin{figure}[tb]
\centerline{\epsfig{file=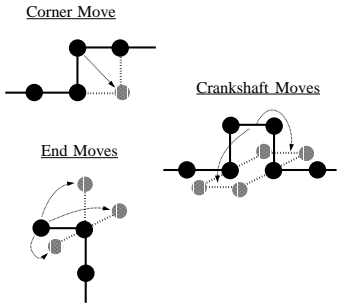,width=.67\figurewidth}}
\caption{The three types of moves used in the
  simulations. The light circles represent the possible lattice
  points to which a given monomer can move, provided that that point
  is not occupied. In the case of the end and crankshaft moves, one of
  the possible moves is picked at random. Note that the corner and
  crankshaft moves are exclusive: A non-end monomer can only make one
  or the other depending on the position of its neighbors along the
  chain.}
\label{fig:mcmoves}
\end{figure}

The move set is shown in figure~\ref{fig:mcmoves} and consists of
local one- or two-monomer moves. These moves are the standard set used
in lattice polymer simulations. They are believed to produce
reasonably realistic dynamics (see
references~\onlinecite{Verdier62,Hilhorst75,Kremer88,Gurler83} for
details). For thermodynamics calculations, it is not necessary to use
a move set with this property, and other move sets have been
used.\cite{OToole92,OToole93} There are however, two potential
problems with realistic dynamic move sets: ergodicity and glassy
behavior. If one is interested only in kinetics, then these are not
really problems but rather properties of the model. For thermodynamic
calculations, inaccessibility is not a problem either, since we can
consider the definition of the model to include only the states
accessible by the move set specified (hence it will be by definition
ergodic). Therefore, as long as the minimum energy cube state is
accessible from an unfolded chain, there will be no problems.

The glass transition presents a more difficult problem for
thermodynamic calculations. At low temperatures, the dynamics of the
system slow down substantially.  In particular, the correlation times
become quite large and it takes longer to explore conformational
space. This gives rise to two errors in the Monte Carlo calculations.
First, it takes a long time for the system to relax, making it
difficult to get the system into thermal equilibrium.  This
contributes a systematic error to the results and it can give rise to
the hysteresis effect seen in the previous
studies.\cite{OToole92,Note01} Second, because of the long
autocorrelation times subsequent samples are no longer statistically
independent of each other. This has the effect of increasing the
variance of any observable (and the corresponding statistical error).
The actual variance is given by:
\begin{equation}
  \sigma^2_{\rm actual} = (1+2\frac{\tau_{\rm ac}}{\tau_{\rm
      samp}})\sigma^2
\end{equation}
where $\sigma^2$ is the usual variance calculated from the samples.
$\tau_{\rm ac}$ is the integrated autocorrelation time and $\tau_{\rm
  samp}$ is the number of time steps between samples (both measured in
units of Monte Carlo steps).\cite{Muller73,Madras88} As $\tau_{\rm
  ac}$ increases, longer simulations are necessary to get
statistically reasonable results.  That is, in a simulation of $N$
steps there will be $N/\tau_{\rm ac}$ ``effectively independent
samples.''\cite{Madras88} At low temperatures the time required to
obtain enough independent samples becomes enormous.  One solution is
to use a different move set, such as the Rosenbluth chain growth
algorithm.\cite{Rosenbluth55,OToole92,OToole93} The auto-correlation
time for this move set does not increase as rapidly, allowing
simulations at lower temperatures. Our solution instead is to run the
simulations well above the glass transition temperature, and then to
use the histogram method to extrapolate to lower temperatures.

\subsection{Techniques for calculating thermodynamics quantities}

For short enough chains (usually in two-dimensions) it is possible to
enumerate all lattice conformations and thereby calculate the
partition function for the system along with any other quantity of
interest.\cite{Lau89} For the three-dimensional 27 monomer chain, it
is not practical\cite{Note02} to enumerate all conformations but it is
possible to enumerate all of the maximally compact (cube)
conformations. One could then approximate the partition function by
just summing over the cube states. Previous studies have used this
method to calculate thermodynamics quantities of this system. It was
hoped that at low temperatures this approximation might be reasonable.
We show later that, although it can give rough qualitative results,
using only the cube states leads to appreciable errors.

Since exact enumeration is impossible, Monte Carlo sampling is used
for calculating thermodynamic quantities. The usual technique runs the
simulation at a given temperature, collecting samples to determine
thermal averages. The process is repeated for several temperatures to
get the averages as a function of temperature. There are several
drawbacks to the standard Monte Carlo procedure. Calculating extensive
variables like the free-energy or entropy is difficult. Also, if one
wants to find peaks or zeros of a given quantity (like the specific
heat peak, to identify the transition temperature), one must scan over
a range of temperatures to located the critical value.

It is possible to extract more information from a single Monte Carlo
run than just the thermal averages at the temperature the simulation
was performed. The technique is called the histogram method or density
of states method. It has a long history and it has been recently
rediscovered by a variety of
authors.\cite{Ferrenberg88,Ferrenberg89,Alves90,Huang91,Bouzida92,Kumar92}
The actual Monte Carlo sampling algorithm itself is unchanged. But
instead of just calculating thermal averages, one keeps track of the
number of times a specific energy is encountered in the simulation;
i.e., an energy histogram is calculated. This histogram, $h(E,T)$,
measures the probability of energy $E$ occuring at temperature $T$. It
is equal to the thermal average of the density of states:
\begin{equation}
  h(E,T') = \frac{n(E)e^{-E/T'}}{Z(T')},
\end{equation}
where $Z(T')$ is the partition function at temperature $T'$ is
\begin{equation}
  Z(T')=\sum_E n(E)e^{-E/T'},
\label{eq:partition}
\end{equation}
and $n(E)$ is the density of states for energy $E$ (the number of
conformations with energy $E$). The Boltzmann factor $k_b$ has been
set equal to 1 and $T'$ is the temperature of the simulation. One
now has the density of states up to a multiplicative factor:
\begin{equation}
  n(E) = h(E,T')e^{E/T'}Z(T'),
\label{eq:dos}
\end{equation}
where $Z(T')$ is the unknown multiplicative constant.  For intensive
quantities, thermal averages are calculated using:
\begin{eqnarray}
  \left\langle{\cal O}\right\rangle(T) &=& \frac{\sum_E{\cal
        O}(E)n(E)e^{-E/T}}{\sum_E n(E)e^{-E/T}}\nonumber\\
    &=&\frac{\sum_E{\cal O}(E)h(E,T')e^{-E/T+E/T'}}{\sum_E
      h(E,T')e^{-E/T+E/T'}}.
\label{eq:thermal}
\end{eqnarray}
Note, $Z(T')$ cancels out of the above expression.

If one is interested in calculating extensive quantities like the free
energy or the entropy, it becomes necessary to determine the constant
$Z(T')$.\cite{Note03} For our system, it is possible to calculate this
constant and therefore obtain the density of states. To determine this
constant, all we need is the multiplicity of any energy state. For
example, the sequences we study have a non-degenerate ground state.
This means $n(E_{\it gs})=1$, where $E_{\it gs}$ is the energy of the
lowest energy cube.  $Z(T')$ can then be determined and the free
energy is then calculated using $F=-T\log Z$ and
equation~\ref{eq:partition}.

\begin{figure}[b]
\centerline{\epsfig{file=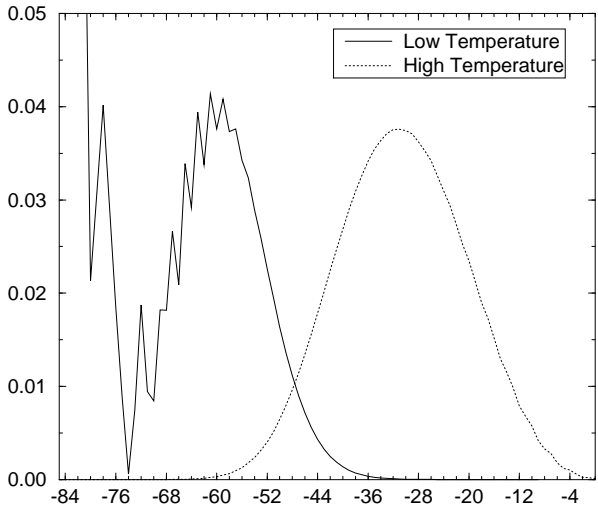,width=\figurewidth}}
\caption{Energy histograms taken at high ($T=3.15$) and low ($T=1.41$)
  temperatures. Note that for each histogram there are regions of
  energy that are not sampled at all. In particular at high
  temperatures the ground state is not probed while at low
  temperatures the high energy (unfolded) conformations are not
  probed.}
\label{fig:histogram}
\end{figure}

There is a limit to the temperature range over which
equation~\ref{eq:thermal} is valid. For temperatures too far from the
simulation temperature, the errors in the density of states calculated
from equation~\ref{eq:dos} become significant. At any given
temperature, the system is only sampling a given region of phase
space.\cite{Note04} For example, figure~\ref{fig:histogram} shows
energy histograms at a high and a low temperatures.
For the high $T$ simulations, the ground state is never probed, and
likewise for the low $T$ some high energy states are never reached.
Consequently, the density of states will be incorrect for regions not
sampled properly (in fact it equals zero for regions that are never
sampled).  This limits the temperature range we can extrapolate from
any simulation.  One needs to monitor the errors in the density of
states. A solution to this problem is the multiple histogram
method.\cite{Ferrenberg89} The idea is to use several different
simulations and patch the histograms together. Although there are some
subtleties to this technique, it can be powerful.

For the 27 monomer long polymer used here, a single histogram gives
adequate results over the range of temperatures of interest. The
reason is that the width of the energy histograms in general scale as
$1/\sqrt{N}$, where $N$ is the system size.  As we shall show shortly,
the system is small enough to insure that the histograms are broad and
a large region of phase space is sampled at any given temperature.

\section{RESULTS AND DISCUSSION}

\begin{figure}[tb]
\centerline{\epsfig{file=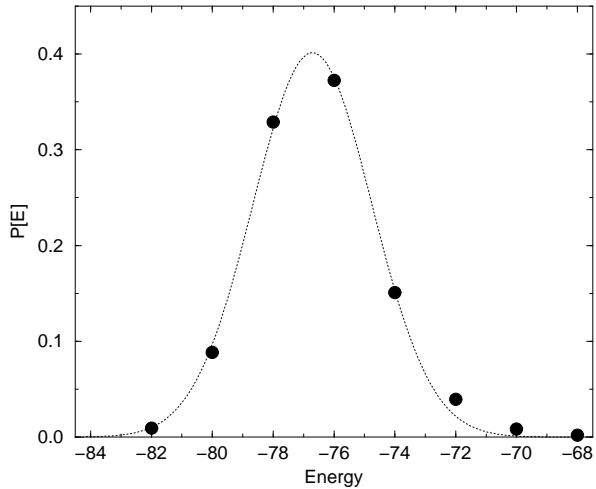,width=\figurewidth}}
\caption{Ground state energy histogram for non-degenerate sequences. 13,563
  sequences with a fix ratio of monomer types (14:13) were generated
  at random. The energy and multiplicity of the minimum energy cube
  was determined from exhaustive enumeration of the cube
  conformations.  1,061 sequences (7.8\%) were found to have unique
  ground states. The histogram of ground state energies from these
  sequences is plotted (solid circles). The dotted line is a lest
  squares fit of a gaussian to the data. Note, we did not find any
  minimal energy (-84) sequences in our random sample. The total
  number of possible sequences is ${27\choose 14}=20,058,300$.  }
\label{fig:energydist}
\end{figure}

Several Monte Carlo runs were performed over a range of temperatures.
Six sequences were used all with a fixed ratio of monomer types (14 to
13).  Table~\ref{tab:seq} lists the sequences and the energy of their
native states.
\begin{table}[tbp]
\begin{center}
\begin{tabular}{lcr}
  \hline\hline
Run & Sequence & $E_{\rm min}$ \\ \hline
002 & {\footnotesize\tt ABABBBBBABBABABAAABBAAAAAAB} & -84 \\
004 & {\footnotesize\tt AABAABAABBABAAABABBABABABBB} & -84 \\
005 & {\footnotesize\tt AABAABAABBABBAABABBABABABBB} & -82 \\
006 & {\footnotesize\tt AABABBABAABBABAAAABABAABBBB} & -80 \\
007 & {\footnotesize\tt ABBABBABABABAABABABABBBABAA} & -80 \\
013 & {\footnotesize\tt ABBBABBABAABBBAAABBABAABABA} & -76 \\ \hline\hline
\end{tabular}
\end{center}
\caption{The various sequences used in this paper. The last four (005,
  006, 007, 013) were generated at random. Sequence 002 was optimized
  by Shakhnovich (Ref.~\protect\onlinecite{Shakhnovich93}). Sequence
  004 is a single monomer mutation of 005 ($B_{13}\rightarrow A$).
  Both 002 and 004 have the lowest energies possible for the potential
  used and have native states that are completely unfrustrated.
  $E_{\rm min}$ is the energy of the native states. These same six
  sequences were studied in our previous work
  (Ref.~\protect\onlinecite{Socci94b}) which examined the kinetics of
  folding.}
\label{tab:seq}
\end{table}
To get some idea of what a typical (i.e. randomly chosen) sequence is
and how it compares to the sequences used in this study, we generated
over 10,000 random sequences (with a 14:13 ratio of monomer types).
Approximately 1,000 sequences had unique ground states.
Figure~\ref{fig:energydist} shows a histogram of the ground
state energies for these sequences.
The distribution is roughly gaussian.  The most probably energy is
approximately -76. One sequence examined (sequence 013) has a typical
ground state energy for random sequences.  We did not generate any of
the minimum energy (-84) sequences at random. The two that we used
where both designed: one using a Monte Carlo
algorithm\cite{Shakhnovich93} and the other by mutating a -82 energy
sequence. The thermodynamic averages of several quantities (average
energy, contacts, native contacts, specific heat, etc.) were
calculated at each of the simulation temperatures.  In addition,
histograms of the number of like versus unlike contacts ($N_l$ and
$N_u$) were calculated. We chose to histogram these variables instead
of the energy directly because these histograms can be used to
extrapolate not only other temperatures but other parameter values
($E_l$ and $E_u$).

\begin{figure}[tb]
\centerline{\epsfig{file=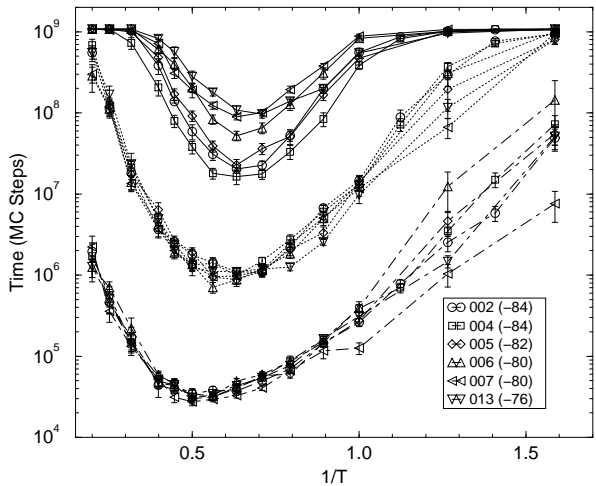,width=\figurewidth}}
\caption{Folding and collapse times versus inverse temperature.
  Time is in Monte Carlo steps. The solid lines represent the mean
  folding time. The middle dotted lines represent the mean compaction
  time to any cube. The bottom lines represent the mean compaction
  time to a partially compact conformation with 25 (out of 28)
  contacts. Error bars are the standard deviation of the mean. Note,
  simulations were run for a set amount of time $\tau_{\rm max}$. For
  high and low temperatures some runs were not able to find the folded
  (or compact) state in this time. In these cases $\tau_{\rm max}$ was
  averaged in, so the times shown are lower bounds to the actual mean
  first-passage times.}
\label{fig:arren1}
\end{figure}

Figure~\ref{fig:arren1} shows the folding time and collapse times as a
function of the inverse temperature. The two different collapse times
are the time to find the first cube (i.e., a maximally compact state)
and the time to form the first 25 (out of 28) contacts. For high
temperatures, the collapse times are sequence independent ({\em
  self-averaging}).  The folding times are sequence dependent.
Similarly, the collapse times below the glass transition are also
sequence dependent. The kinetic glass transition temperature was
defined in our previous work\cite{Socci94b} as the temperature at
which the folding time is half way between its minimum and maximum
value (the maximum being determined by the time limit on the
simulation and chosen to be much longer than the fastest folding
time). Both the folding and collapse time show non-Arrhenius behavior
at high temperatures.

\subsection{Histograms: First vs. second order transitions}

\begin{figure}[ptb]
    \epsfig{file=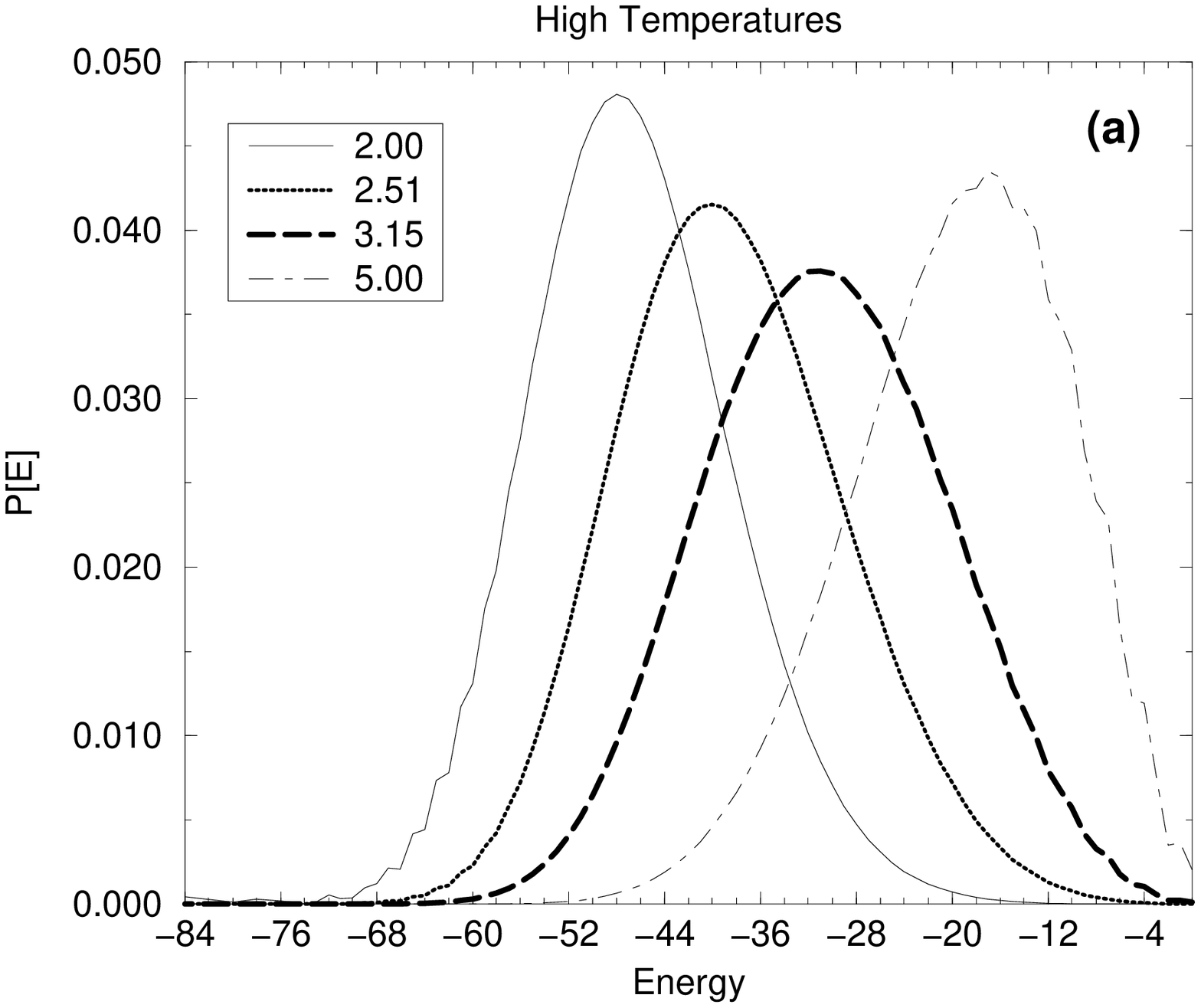,width=\dblfigurewidth}
    \vskip 2\baselineskip
    \epsfig{file=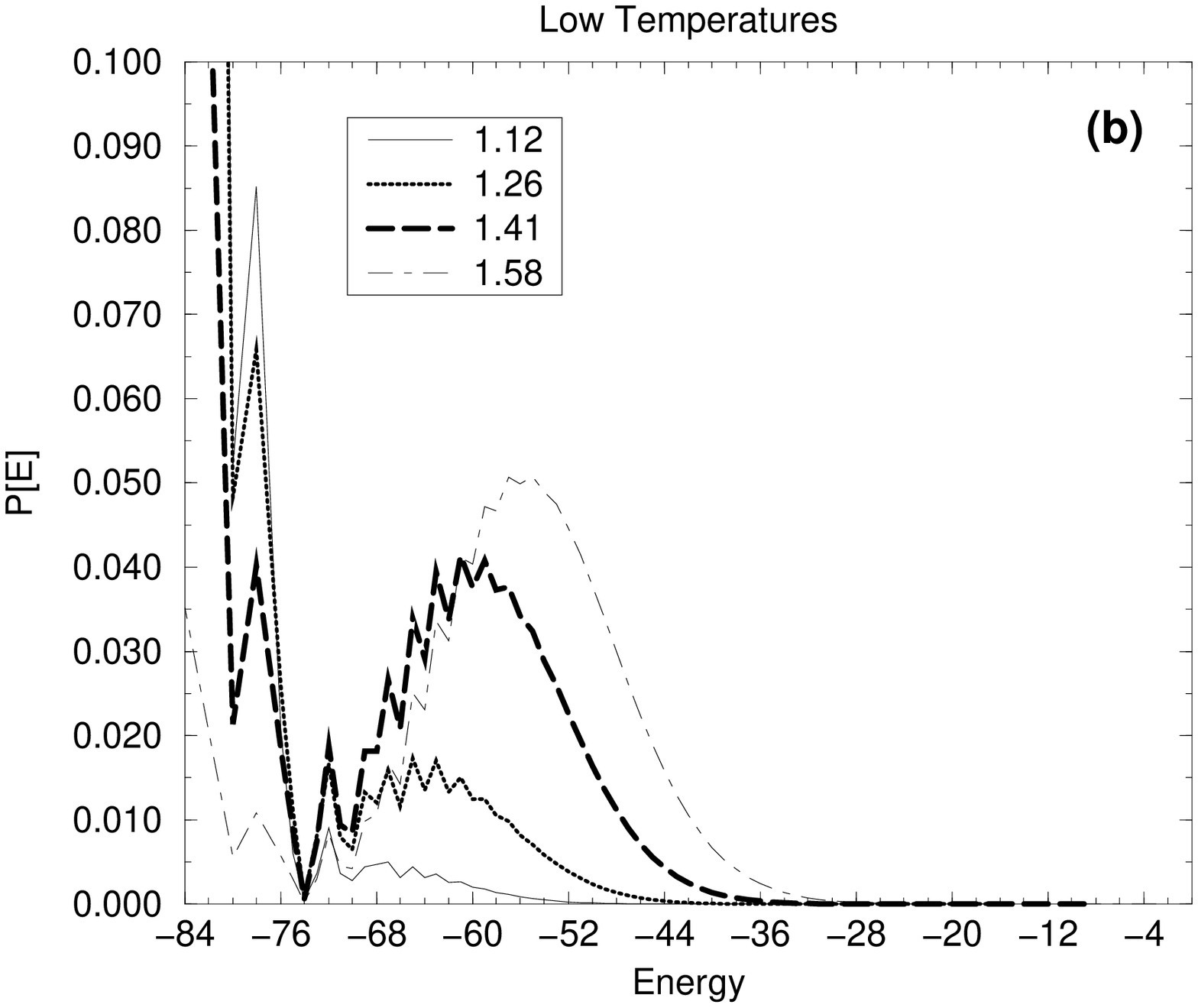,width=\dblfigurewidth}
  \caption{Histograms of energy for several temperatures. The sequence
    is number 002. For high temperatures (a) the plots have one peak
    which moves to lower energies as the temperature decreases. For
    low temperatures (b) the plots are bimodal.}
\label{fig:ehist}
\end{figure}

\begin{figure}[ptb]
  \epsfig{file=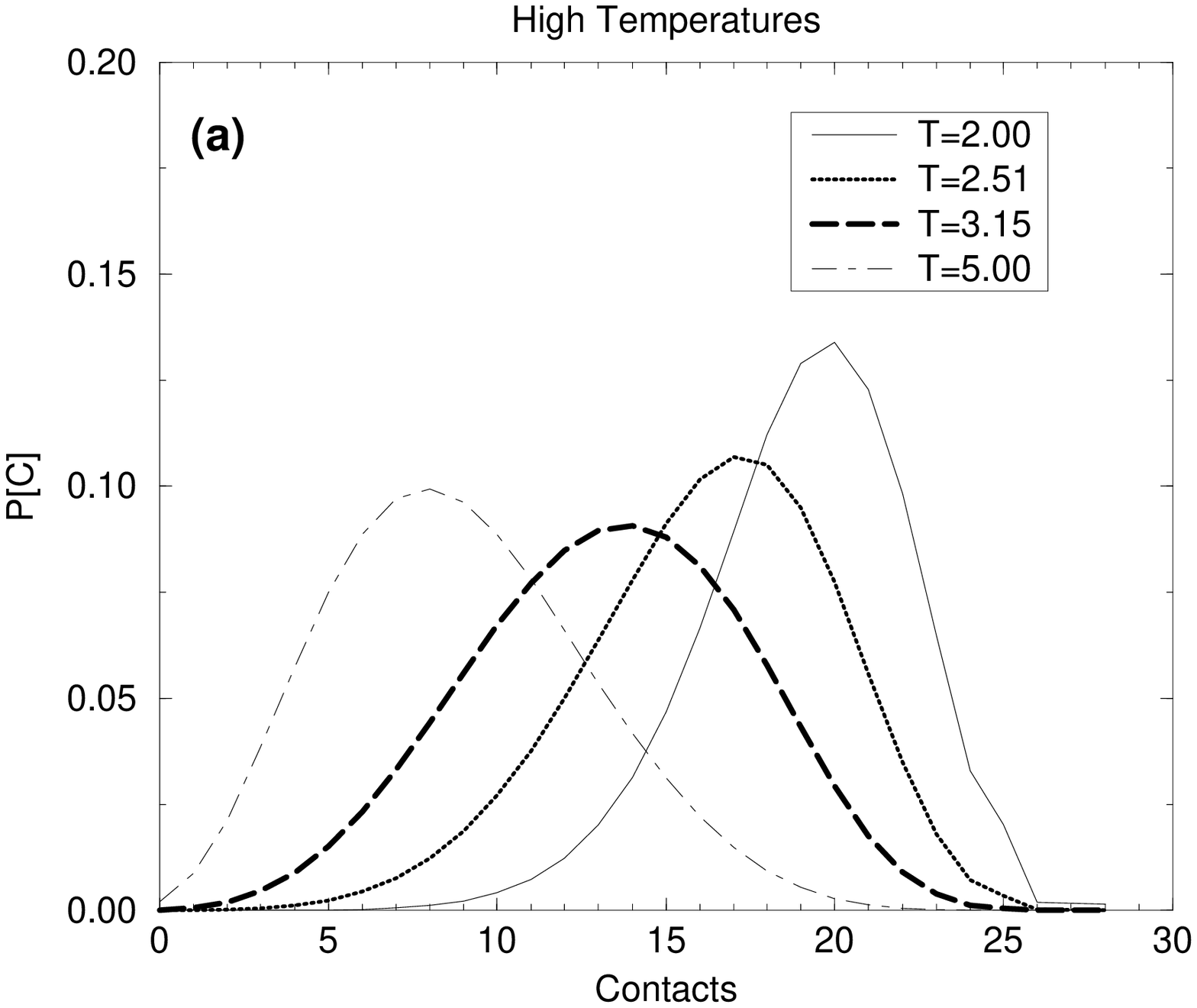,width=\dblfigurewidth}
  \vskip 2\baselineskip
  \epsfig{file=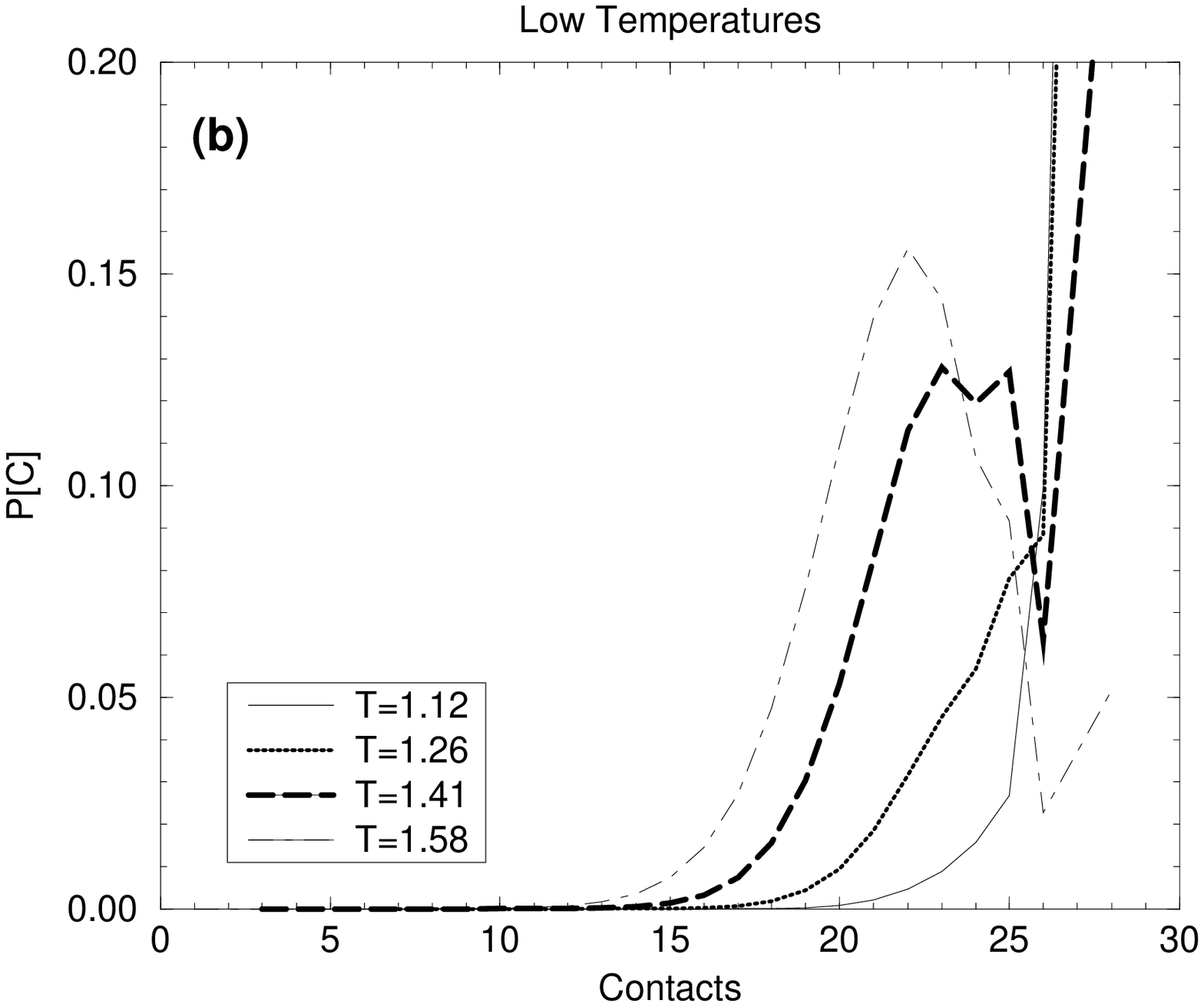,width=\dblfigurewidth}
\caption{Histograms of contacts for several temperatures, again for
  sequence 002. The behavior of the plots is similar to that of the
  energy histograms (see fig.~\protect\ref{fig:ehist}). As the
  temperature is lowered a single peak moves from a low to a high
  number of contacts until it reaches roughly 20 (a). At this point a
  second peak forms at 28 contacts and we see a first order-like
  transition between the two (b).}
\label{fig:chist}
\end{figure}

Before using the histograms to calculate the density of states, we
examined them to determine how much of phase space is sampled at
different temperatures. This (as mentioned above) determines how far
we can extrapolate from the simulation temperature using the histogram
technique. Figure~\ref{fig:ehist} shows the energy histogram for
sequence 002 for several temperatures.
Because of the small size of our system, they are all rather broad,
with widths of roughly 30 energy units. Examining the behavior of the
curves as a function of temperature, we see the both first and second
order-like behavior of the system. At high temperatures (between 5 and
2, see fig.~\ref{fig:ehist}a) a single energy peak moves steadily to
lower values. This is what would be expected from a second order-like
transition. At lower temperatures (fig.~\ref{fig:ehist}b) the plots
now have a bimodal distribution and as the temperature is decreased
there is a shift from one peak to the other. This is characteristic of
a first order-like transition. If we examine the histograms as a
function of the number of contacts, the same behavior is seen
(fig.~\ref{fig:chist}).
At high temperatures the plots are unimodal
and shift to larger number of contacts (higher compactness) as the
temperature is lowered. This continues until the maximum reaches
roughly 20 contacts around $T=2$. At lower temperatures this peak
remains fixed at about 20 contacts and another peak forms at 28
contacts. Because this peak at 28 contacts occurs at low temperatures
and when there is a peak at -84 in the energy histograms, we expect
that it is due to occupation of the native state and the few other low
energy cubes. As the the temperature is decreased there is a shift in
population between the two peaks. This is consistent with the idea
that there are two thermodynamic transitions: a collapse to compact
structures and a folding transition to the native state. The collapse
transition occurs at a higher temperature and is second order-like.
The folding transition is first order-like.

Since the histograms are broad enough we used the single histogram
technique in the subsequent calculations. Because we are interested
mainly in the properties of the ground state, we chose a temperature
which is low enough for the ground state to be sufficiently populated
and yet high enough so that we sample as much of the conformation
space as possible. At a temperature of 1.58 there is a sizeable peak
at the ground state and substantial sampling of the higher energy
conformations. Kinetically it turns out that $T=1.58$ the chains fold
most rapidly (i.e., the mean first-passage time to the folded state is
smallest). So we expect at this temperature we are moving most rapidly
through the compact conformations. Since this temperature is far
enough away from the kinetic glass temperature (which was previously
measured to be approximately 1 for this system) we do not have to
worry about the problem of long relaxation times which would make it
difficult to equilibrate and would require a long sampling time to
reduce statistical errors. We calculated the energy autocorrelation
function:
\begin{equation}
  C_{E}(t) = \frac{\left\langle E_\tau E_{\tau+t}\right\rangle
         -\left\langle E\right\rangle^2}{%
         \left\langle E^2\right\rangle-\left\langle E\right\rangle^2}
\label{eq:autocorr}
\end{equation}
where $E_\tau$ is the energy of the system at time step $\tau$. At
long times this function should have the following form:
\begin{equation}
  C_{E}(t) \sim e^{-t/\tau_{\rm ac}}
\label{eq:tauac}
\end{equation}
where $\tau_{\rm ac}$ is the autocorrelation time.\cite{Madras88} At
the temperature $T=1.58$ we get an autocorrelation time of roughly
500,000 Monte Carlo steps. For all our thermodynamic simulations we
equilibrated our system for $20\tau_{\rm ac}$ and ran for
$1.08\times10^9$ steps (2000 times $\tau_{\rm ac}$ which gives roughly
2000 independent samples).

\subsection{Density of states}

\begin{figure}[ptb]
  \epsfig{file=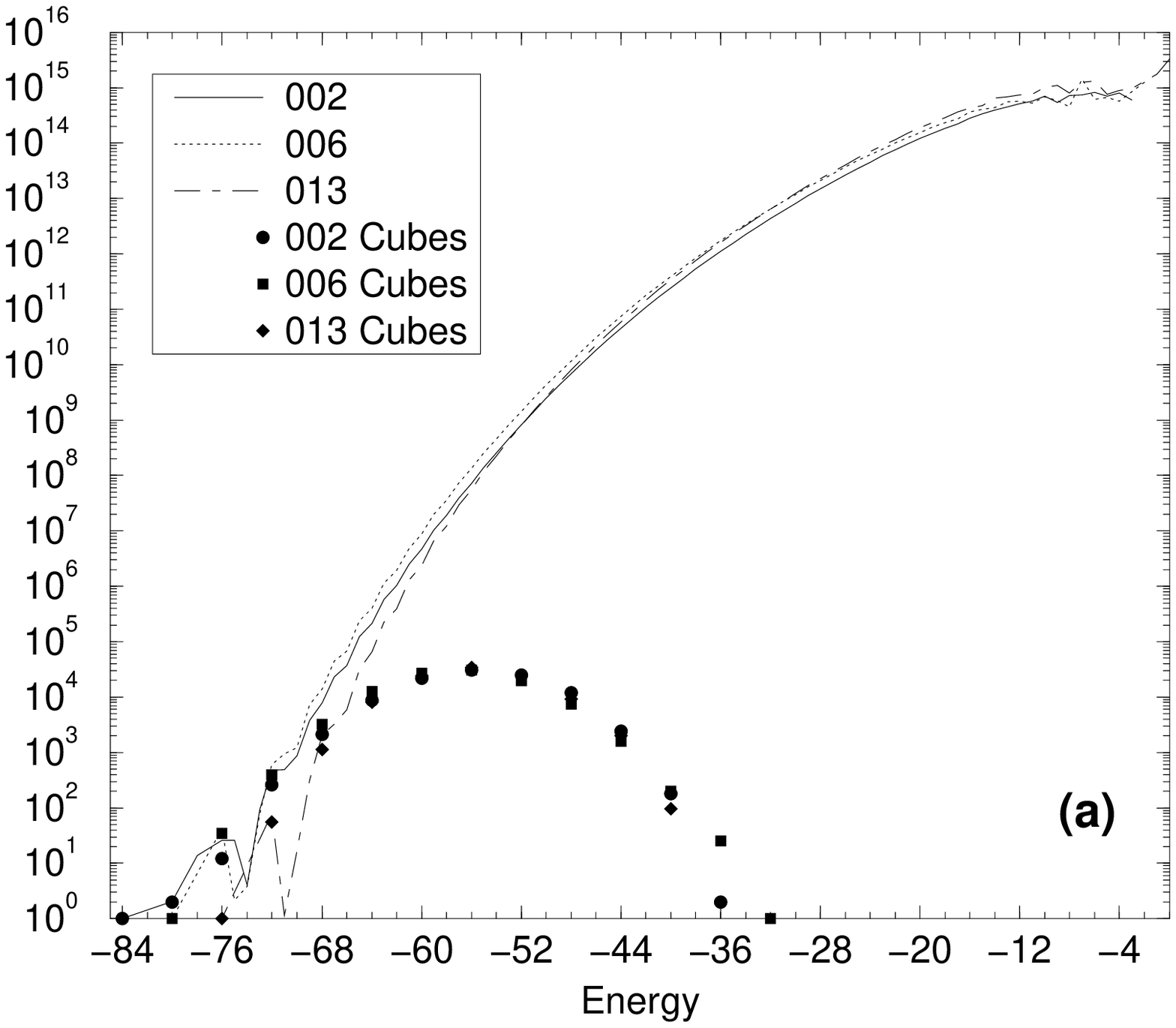,width=\dblfigurewidth}
  \vskip 2\baselineskip
  \epsfig{file=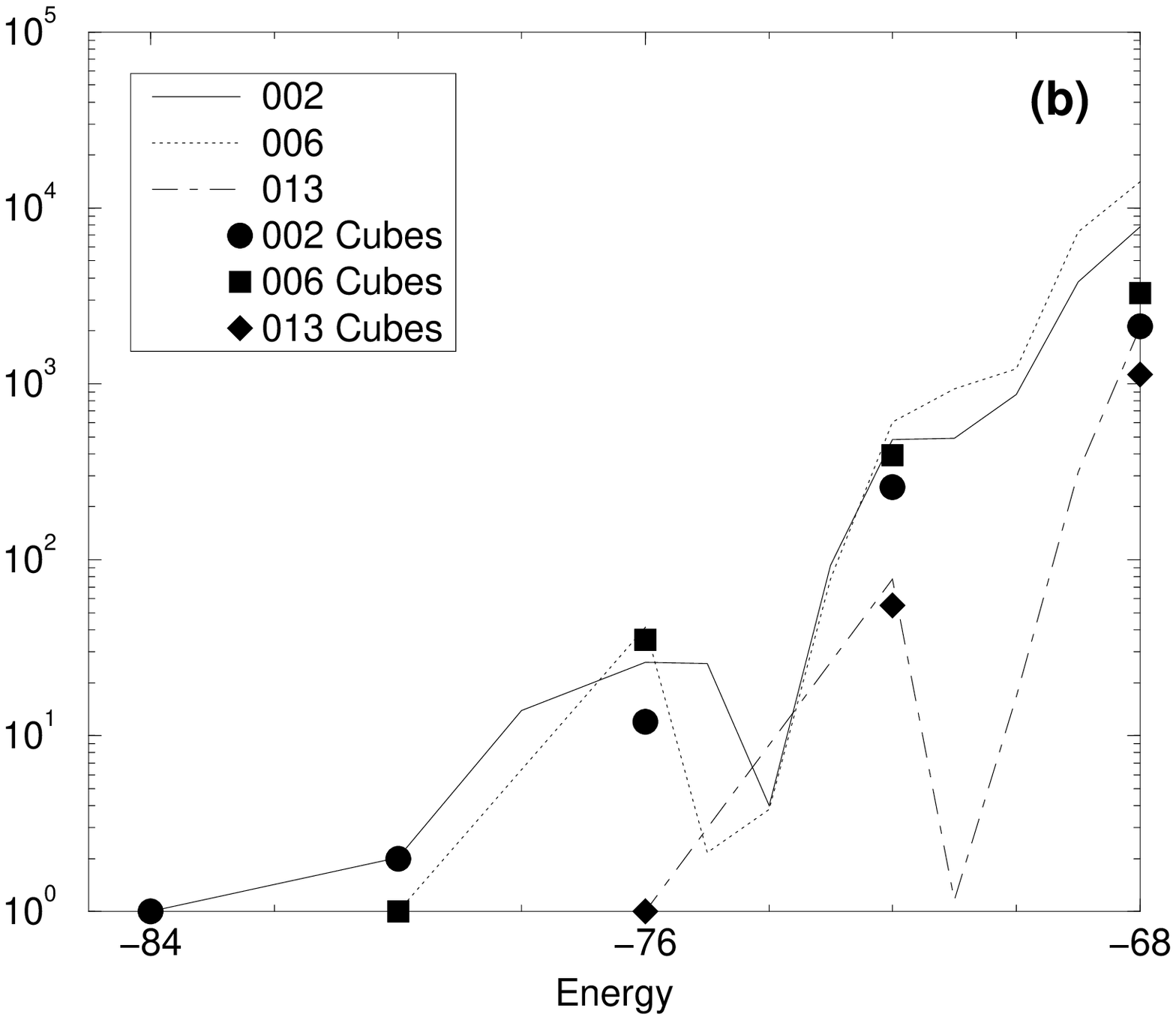,width=\dblfigurewidth}
\caption{Density of states versus energy calculated using the Monte
  Carlo histogram technique. Three different sequences are shown (002,
  006, 013) with different ground state energies. The first plot (a)
  is the full density of states. The lines are from the Monte Carlo
  calculation. The points show the density of states for just the cube
  conformations (determined by exact enumeration). The second plot (b)
  is a blow-up showing the low energy region of the first. At high
  energies the densities of states are sequence independent while at
  low energies they are strongly sequence dependent.}
\label{fig:dosE}
\end{figure}

Using the histograms from the Monte Carlo simulations and
equation~\ref{eq:dos} the density of states for the various sequences
is calculated. Figure~\ref{fig:dosE} shows the densities for three
sequences with low, medium and high values for $E_{\rm min}$ (002,
006, 013).
For comparison the cube spectrum, determined by exact
enumeration, has also been plotted. At low energies the the density of
states is different for each sequences. In particular, sequence 002
has more low energy conformations that are not cubes as compared to
006 and 013.  All three sequences have a notch in their density, but
the gap between the notch and the folded state is larger for sequence
002 than for the others.

At energies below $-60$ the densities of states for the three
sequences are roughly the same. This part of the spectrum is {\em
  self-averaging} as expected since it should depend on only the ratio
of monomer types. At very high energy there is considerable scatter in
the plots. This is due to the poor sampling of this area of
conformational space. In particular, the curves in
figure~\ref{fig:dosE} for sequence 002 and 006 do not even extend to
zero energy, indicating that these conformations are not sampled at
all. However, we expect that for low temperature thermodynamic
calculations this will not pose problems.

\begin{figure}[tb]
\centerline{\epsfig{file=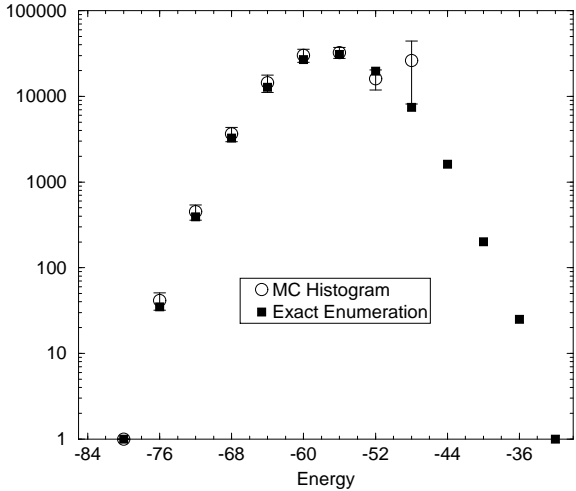,width=\figurewidth}}
\caption{Comparison of the cube spectrum from exact enumeration and
  from the Monte Carlo histogram calculation. Error bars are the
  standard error of the mean from several Monte Carlo runs. The
  temperature of the simulations was 1.58. At this temperature the
  average energy is $-55.5$ and the percent of population in the
  ground state is $0.2\%$.}
\label{fig:cubevsMC}
\end{figure}

As a simple check of the accuracy of the Monte Carlo histogram
technique in this system, we compare the exact cube spectrum (from
enumeration of all cubes) to the cube spectrum calculated from the
histogram data.  Remember that there is an unknown normalization
factor, which we determined by setting the density of states for the
lowest energy cube equal to 1. Figure~\ref{fig:cubevsMC} shows the
comparisons.
For cubes up to an energy of $-52$ there is excellent
agreement between the histogram calculation and the exact answer. At
high energies we see the same sampling problem; cubes with energy
greater than $-50$ are not sampled at all, since they make up a
negligible fraction of the conformations at these energies. However,
for low temperature calculations the errors should be negligible.

\subsection{Computing thermodynamic quantities}

\begin{figure}[tb]
\centerline{\epsfig{file=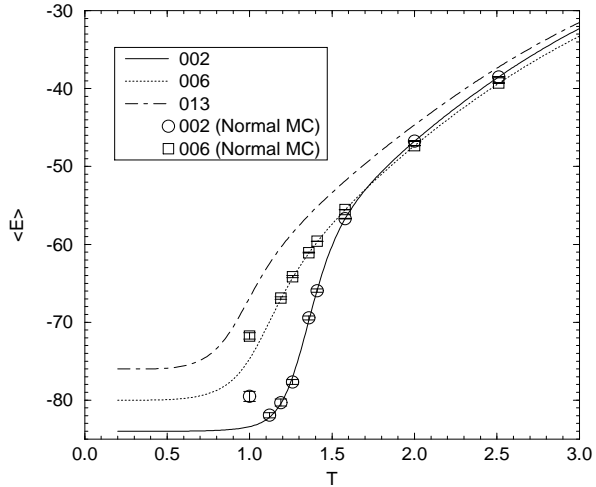,width=\figurewidth}}
\caption{Average energy versus temperature for three different
  sequences (002, 006 and 013). The lines are calculated using the
  histogram technique from simulations at $T=1.58$; the points are
  from normal Monte Carlo simulations (i.e., they were calculated from
  the usual averaging technique at several different temperatures).
  For most temperatures there is excellent agreement between the two.
  As we approach the glass temperature the normal Monte Carlo
  technique starts to deviate due to the divergence of the relaxation
  (equilibration) time of the system.}
\label{fig:eavg}
\end{figure}

\begin{figure}[tb]
\centerline{\epsfig{file=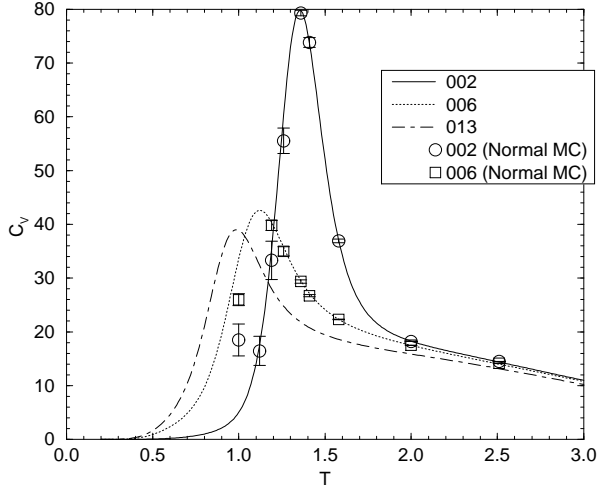,width=\figurewidth}}
\caption{Specific heat versus temperature for sequences 002, 006 and
  013. The lines are calculated using the histogram technique from
  simulations at $T=1.58$; the points are from normal Monte Carlo
  simulations (i.e., they were calculated from the usual averaging
  technique at several temperatures). At the kinetic glass temperature
  ($T=1$) there is substantial error in the normal Monte Carlo
  result.}
\label{fig:cv}
\end{figure}

Figure~\ref{fig:eavg} is a plot of the average energy as a function of
temperature for the same three sequences (002, 006 and 013) whose
densities of states are shown in figure~\ref{fig:dosE}. At high
temperatures ($T>2.5$) all their sequences have the roughly the same
average energy. At lower temperatures the sequences are no longer
self-averaging. The two sequences with the higher energy folded states
(006 and 013) have a fairly broad transition while the low energy
sequences (002) have a comparatively sharper transition. A similar
result is seen in the specific heat, which is plotted in
figure~\ref{fig:cv}.
Sequence 002 has a much sharper and higher
specific heat peak which occurs at a higher temperature. The other
sequences have broader smaller peaks. The peak in the specific heat
occurs at temperatures slightly higher than the folding temperature
(see tab.~\ref{tab:cubevsfull}) and indicates the transition from the
unfolded chain to the collapsed state rather than the transition to
the native state.  At high temperatures the specific heat is sequence
independent.

Also included in figures~\ref{fig:eavg} and~\ref{fig:cv} are data
points calculated using the standard Monte Carlo averaging technique
for comparison. There is excellent agreement between the histogram
curves and the points up until the kinetic glass temperature (which is
at $T\approx1$ for these sequences). At the glass temperature the
usual Monte Carlo technique has a problem with the increasing
autocorrelation time (all simulations were equilibrate and sampled for
the same amount of time). The histogram method, however, allows us to
probe beyond the glass temperature since the low energy states can be
sampled accurately at higher temperatures (see
fig.~\ref{fig:cubevsMC}).

\begin{figure}[ptb]
  \epsfig{file=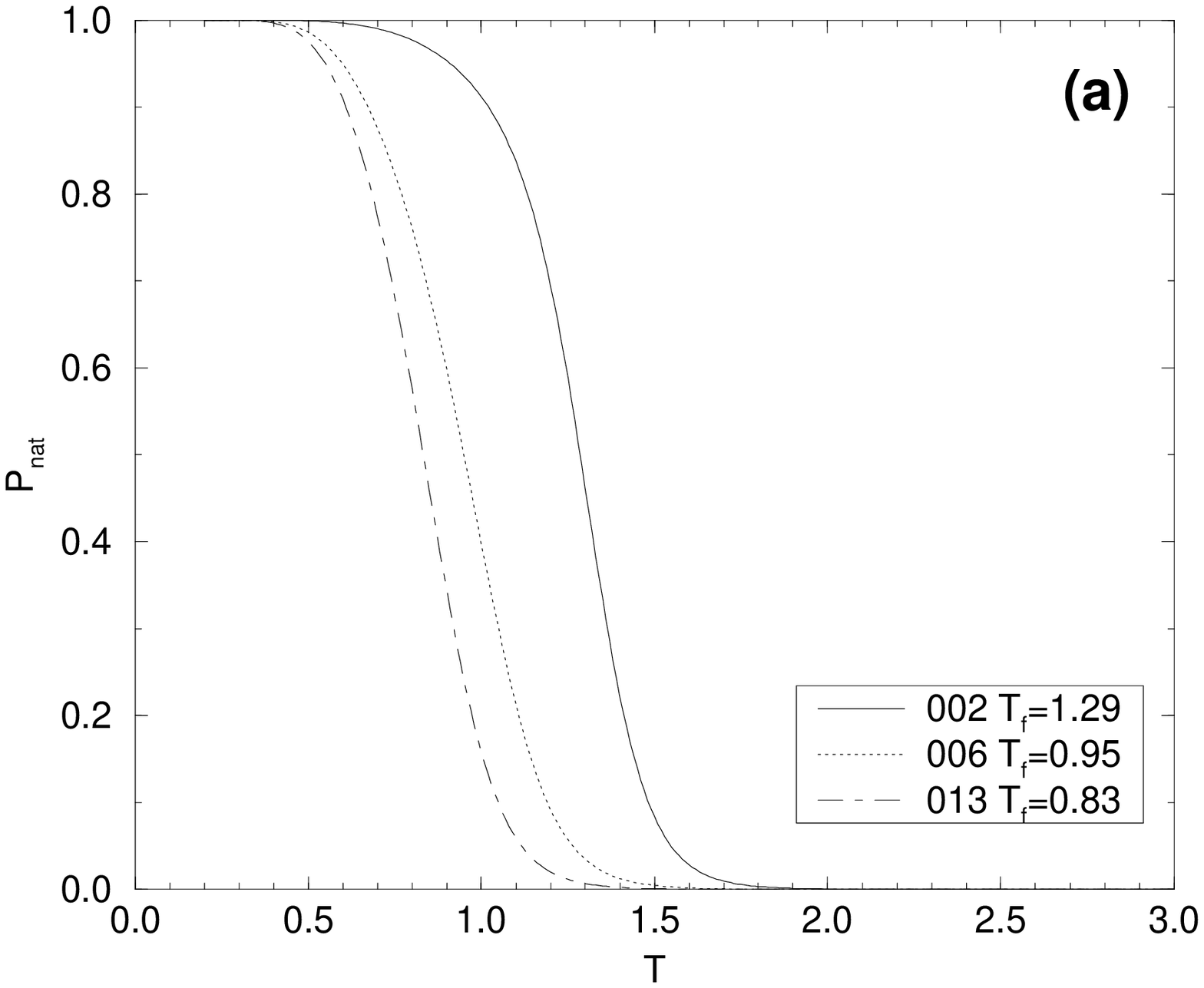,width=\dblfigurewidth}
  \vskip 2\baselineskip
  \epsfig{file=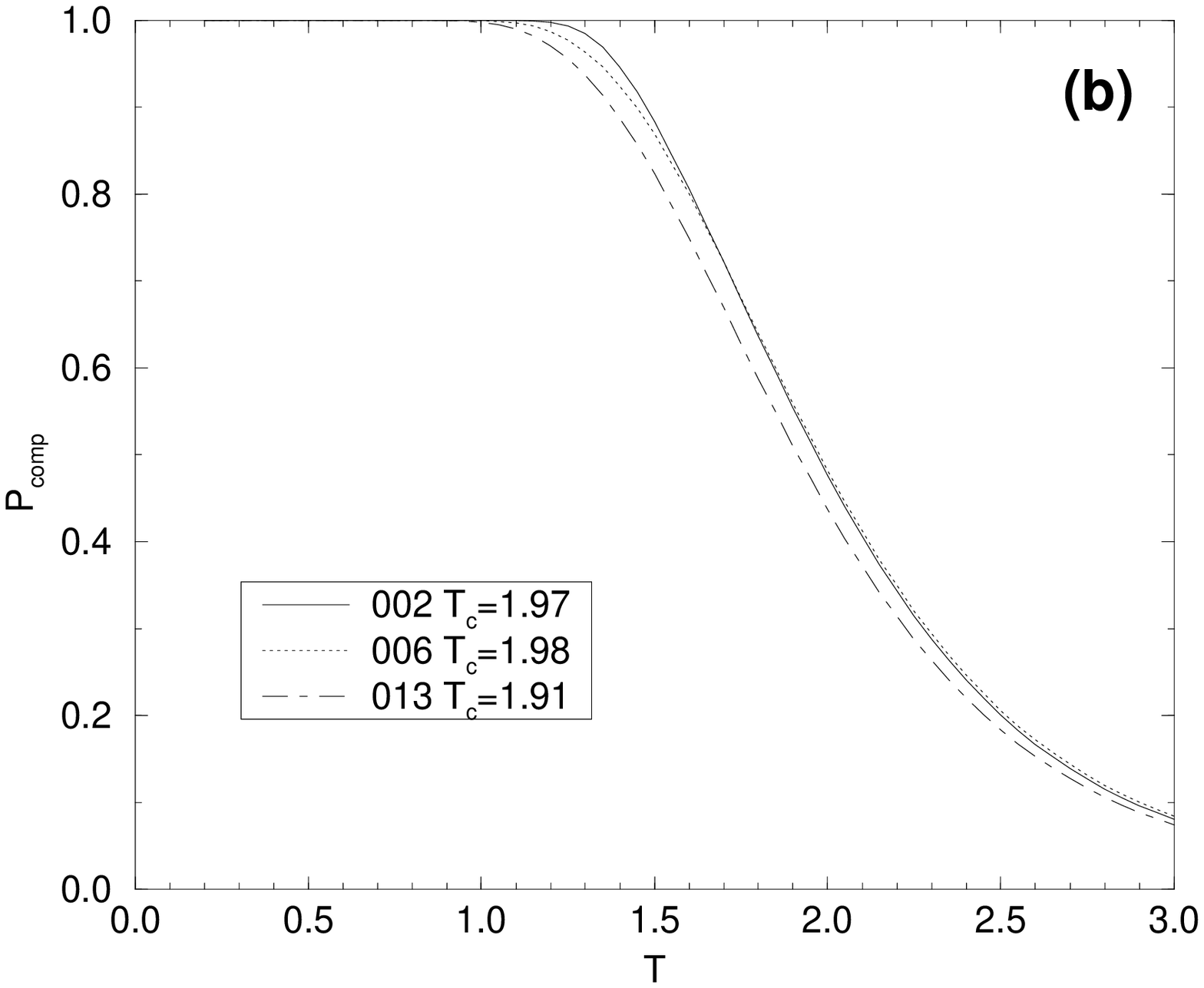,width=\dblfigurewidth}
\caption{
  Folding and collapse transitions for sequences 002, 006 and 013.
  Figure (a) is a plot of the probability to be in the native state
  ($P_{\rm nat}$) versus temperature, while figure (b) plots the
  probability to be compact (20 out of 28 contacts, $P_{c_{20}}$).
  The folding temperature is defined as the temperature at which
  $P_{\rm nat}=\frac{1}{2}$. The collapse temperature is defined
  similarly.  The folding temperature is a sequence-dependent quantity
  while the collapse temperature is roughly sequence-independent
  (self-averaging). We expect the collapse transition to depend on the
  ratio of monomer types (i.e., the over-all drive to compactness) and
  therefore it should not depend on the specific sequence.}
\label{fig:pnat}
\label{fig:pcomp}
\end{figure}

One useful feature of the histogram technique is the ability to
determine extrema and zeros of thermodynamic functions. For example,
the folding temperature $T_f$ is the temperature at which the
population of the native state equals one half:
\begin{equation}
  P_{\rm nat}(T_f) = \frac{e^{-E_{\rm nat}/T_f}}{Z} = \frac{1}{2}.
\label{eq:Tf}
\end{equation}
Once the density of states has been determined, one can numerically
solve for $T_f$ using any standard root-finding
algorithm.\cite{Press86} Figure~\ref{fig:pnat} plots $P_{\rm nat}(T)$
for the three sequences along with the folding temperatures.
Also
shown is a plot of the probability of being semi-compact (which we
define as
structures have 20 or more contacts):
\begin{equation}
  P_{c_{20}}(T) = \frac{\sum\limits_E\sum\limits_{C\ge20}
    n(E,C)e^{-E/T}}{Z},
  \label{eq:pcomp}
\end{equation}
where $n(E,C)$ is the density of states as a function of energy and
contacts. Note that in order to compute this quantity we need to keep
track of histograms as a function of energy and contacts.  Histograms
of just the energy would not have allowed us to sort out the compact
states from the non-compact ones. Similar to $T_f$ the compaction
temperature $T_c$ occurs when the probability of being semi-compact
(20 contacts) equals one half. Figure~\ref{fig:pcomp} shows the
$P_{c_{20}}$ curves along with the values for $T_c$.  We chose 20
contacts because that was the point at which the histograms (see
fig.~\ref{fig:chist}) changed from their second order to first
order-like behavior.  Therefore, we expect that measuring this
quantity will probe the first transition from random coil to globule
(semi-compact) states.

As expected the folding temperature ($T_f$) is sequence dependent.
Also, the transition curves for folding are much sharper than they for
collapse.  The lower the energy of the ground state the higher the
folding temperature for that sequence. In contrast, the compaction
temperature ($T_c$) is almost sequence independent (it varies by only
4\% versus a 43\% difference for $T_f$). One would expect the
compaction temperature to be self-averaging since it should depend on
the average composition of the sequence (which is the same for all
sequences used in this work). At the compaction temperature the native
state occupation ($P_{\rm nat}$) is very small. This is consistent
with the previous observation from the histograms (see
figs.~\ref{fig:ehist} and~\ref{fig:chist}). There are two separate
thermodynamic transition: collapse from a random coil and then folding
to the native state.

It is clear from figure~\ref{fig:dosE} that using just the cubes to
calculate thermodynamics can lead to potentially large errors.
Table~\ref{tab:cubevsfull} compares the folding temperature calculated
using the full density of states with the temperature calculated
solely on the basis of cube states.
\begin{table}[tbp]
\begin{center}
\begin{tabular}{lcrrr}
  \hline\hline
  Run & Full DOS & Cubes & Percent Error \\ \hline
  002 & 1.29(2) & 1.763 & 37.2\%\\
  004 & 1.26(1) & 1.695 & 34.1\%\\
  005 & 1.15(2) & 1.429 & 24.2\%\\
  006 & 0.94(6) & 1.049 & 11.5\%\\
  013 & 0.83(5) & 0.935 & 12.6\%\\ \hline\hline
\end{tabular}
\end{center}
\caption{Comparison of the folding temperature $T_f$ calculated using
  the full density of states (from the histogram method) and just the
  cube states (from exact enumeration). Numbers in parentheses are the
  uncertainty in the last digit. The last column is the percent error
  of the cube-only calculation.}
\label{tab:cubevsfull}
\end{table}
The cube results consistently
over-estimate the folding temperature.  This is not surprising since
many low energy non-cube states are being neglected. These states will
reduce the stability of the native state.  The cube approximation is
better for the non-folding high energy sequences than for the low
energy sequences. Consequently, it fails more seriously for the
sequence we are most interested in, namely the good folding sequences.

\subsection{Combining kinetics and thermodynamics---unfolding time}

Much of the work (both experimental and theoretical) on protein
folding deals with the forward process (unfolded to folded).  However,
studying the reverse process, the unfolding of nascent proteins, may
not only provide a wealth of information, but may also be a great deal
easier. In unfolding simulations, the initial condition is
well-defined (the folded state) and by varying the various parameters
it is fairly easy to induce unfolding. There have been several works
examining unfolding using detailed molecular dynamics
simulations.\cite{Hirst94,LiA94}

\begin{figure}[tb]
\centerline{\epsfig{file=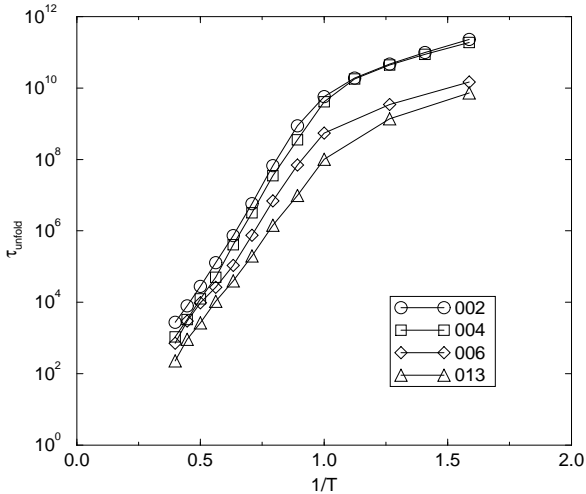,width=\figurewidth}}
\caption{Unfolding times versus $1/T$ for several sequences. The time
  is plotted on a log scale. The linear relationship between time and
  $1/T$ for temperatures greater than $T_g$ is much simpler than the
  behavior of the folding time.}
\label{fig:unfold}
\end{figure}

Using the data previously calculated, we can compute an unfolding time
for our lattice chains. We first make the two-state assumption, namely
that there is an unfolded state ({\sf U}) that is in thermal
equilibrium
with the folded state ({\sf F}):
\begin{equation}
  {\sf F}\rightleftharpoons{\sf U}.
\end{equation}
We can then calculate an unfolding time ($\tau_{u}$) as follows:
\begin{equation}
  \tau_{u}(T)=\frac{\left[{\sf F}\right]}{\left[{\sf
      U}\right]}\,\tau_{f}(T)
  \label{eq:tauu}
\end{equation}
where $\left[{\sf U}\right]$, $\left[{\sf F}\right]$ are the
populations of the unfolded and folded state respectively and
$\tau_{f}(T)$ is the folding time as a function of temperature. The
ratio $\left[{\sf F}\right]/\left[{\sf U}\right]$ is given by $P_{\rm
  nat}(T)/(1-P_{\rm nat}(T))$, using $P_{\rm nat}(T)$ defined by
eq.~\ref{eq:Tf}. Figure~\ref{fig:unfold} shows a plot of the unfolding
time versus $1/T$ for several sequences.
Unlike the folding time (fig.~\ref{fig:arren1}), the unfolding time
has a much simpler behavior. For temperatures above the kinetic glass
temperature ($T_g$), the unfolding times vary almost linearly with
$1/T$ and have slightly different slopes for the various sequences.
The slopes are roughly proportional to $T_f$; sequences 004 and 002
have the steepest slopes and 013 has the shallowest. In contrast to
the folding time, which shows clear non-Arrhenius behavior over this
temperature range, the unfolding time is nearly Arrhenius. We expect
due to the large enthalpic barrier that entropic effects are less
noticeable in unfolding than folding.  At low temperatures the
unfolding time rolls off due to the cutoff in the simulation time
(this is the same roll-off seen in the folding time at low
temperatures; see fig.~\ref{fig:arren1}).

\subsection{Exploring parameter space}

\begin{figure}[tb]
\centering
\makebox[.95\figurewidth]{\epsfig{file=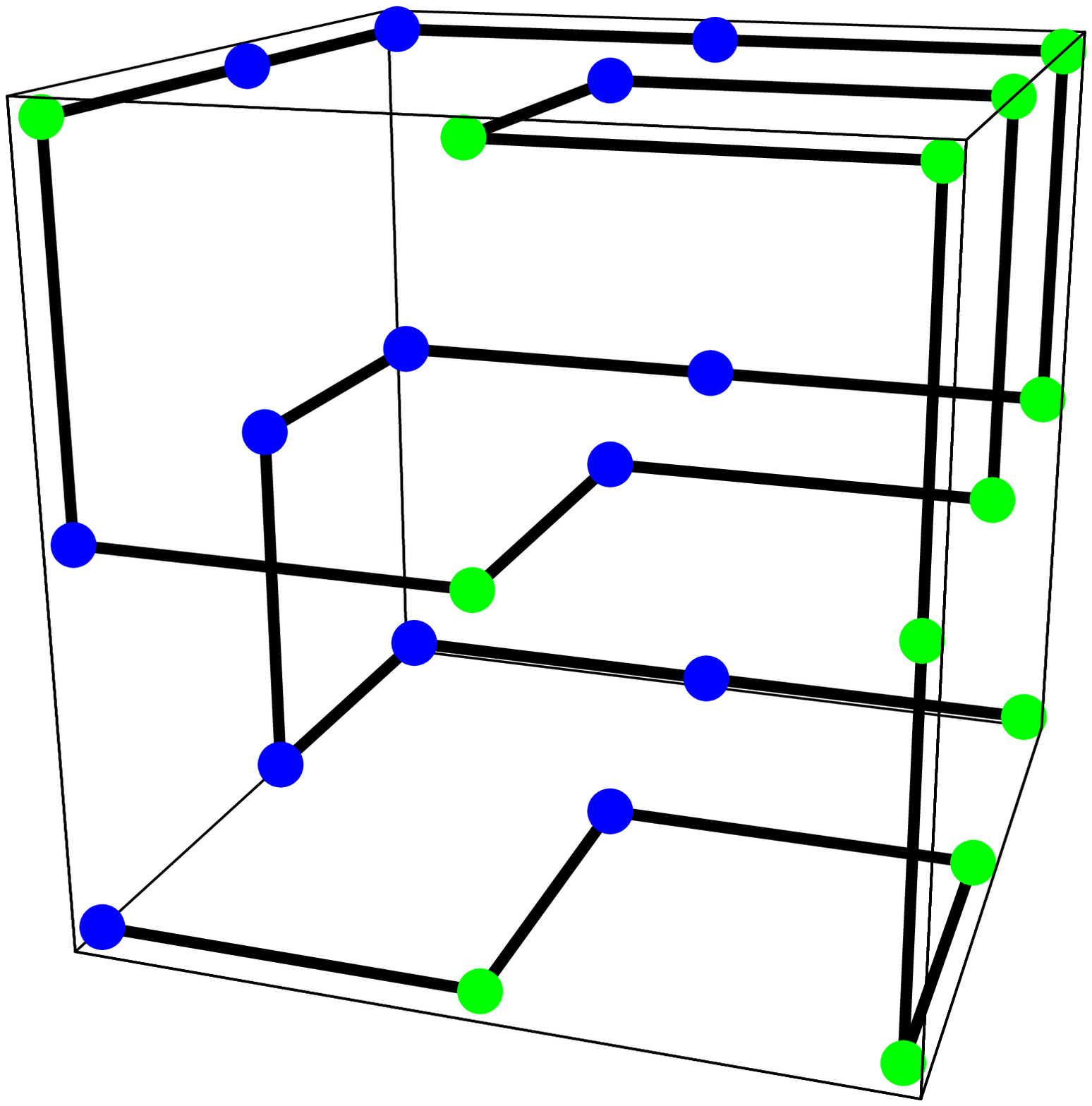,width=.425\figurewidth}%
  \hfill\epsfig{file=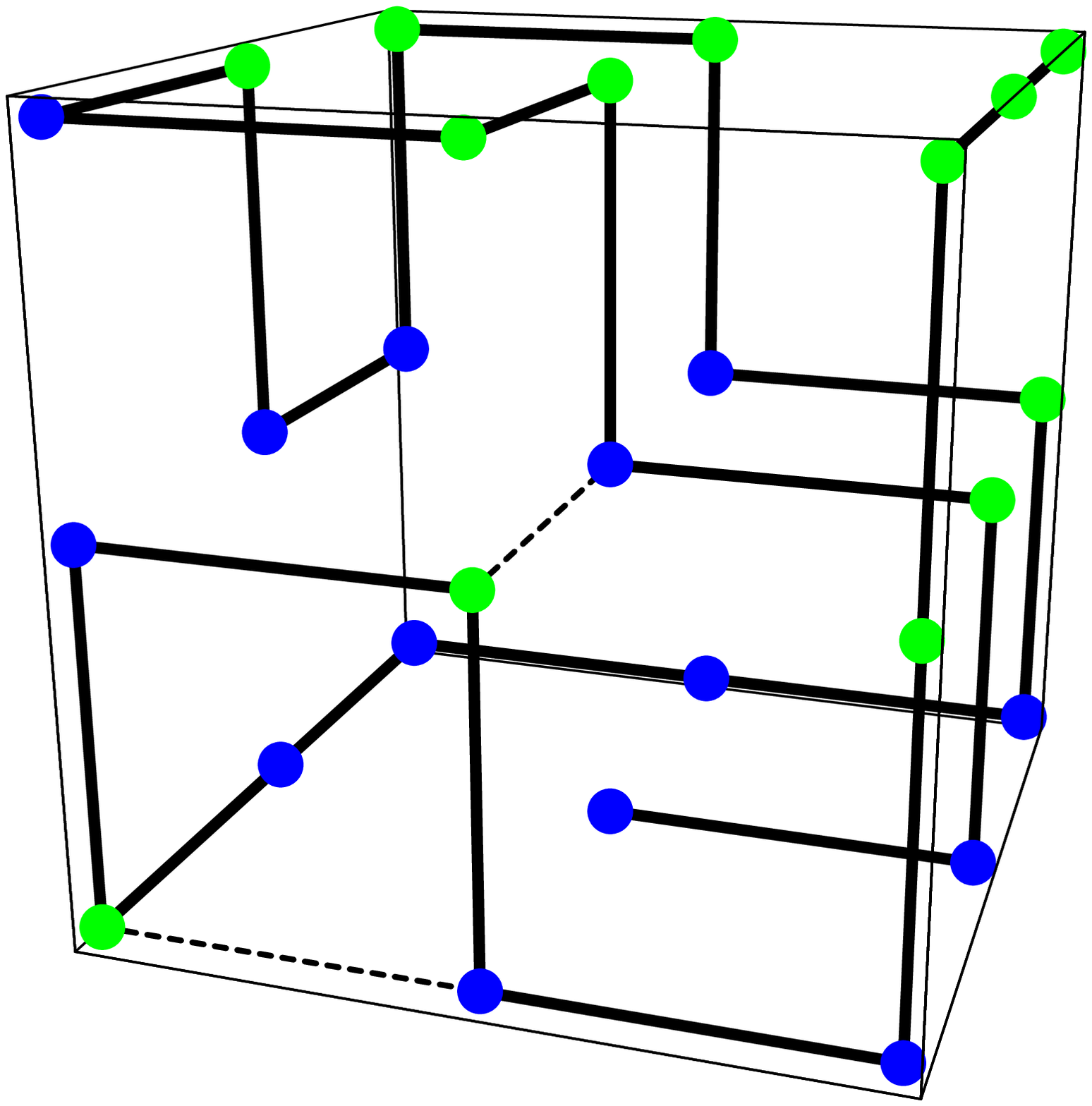,width=.425\figurewidth}}
\caption{The native conformations of sequence 002 (left) and 006
  (right). Note that 006 has 2 ``weak'' contacts (indicated in the
  figure with dotted lines) in its lowest energy state.}
\label{fig:cubes}
\end{figure}

All the results so far have been for simulations with contact energies
$E_{\rm avg}=-2$ and $\Delta=2$ ($E_l=-3$ and $E_u=-1$). These values
were chosen to insure that the ground state (native state) would be a
cube.  For sequences like 002 and 004 there exists a cube (out of the
103,346 possibilities) that has no weak contacts (contacts between
different monomer types). This cube will be the ground state as long
as $E_l=(E_{\rm avg}-\Delta/2)<0$, irrespective of what $E_u$ is
(i.e., $E_u$ can be greater than zero). We call these sequences {\em
  unfrustrated}.  All of the other sequences have at least one weak
contact between unlike monomers even in their lowest energy cube (see
fig.~\ref{fig:cubes}).
These are {\em frustrated} sequences.  For
frustrated sequences, it is not clear that the minimum-energy cube
will be the minimum-energy conformation. There may be some non-cube
conformation with fewer total contacts but more good contacts than the
cube conformation.  For sequence 005, which has 27 good contacts in
its minimum cube conformation, and 006 and 007, which have 26 good
contacts, there are no other conformations that have more good
contacts.  Consequently, the cube conformations will be the ground
states as long as $E_u=(E_{\rm avg}+\Delta/2)<0$.\cite{Note09} For
sequence 013, which has 24 good contacts in its minimum-energy cube,
it is possible that there is some non-cube conformation with more than
24 good contacts.  We can not exhaustively enumerate all the non-cube
conformations but we can state empirically that no conformation with a
lower energy was found in any of the Monte Carlo runs (see
note~\onlinecite{Note09}).

\begin{figure}[tb]
\centerline{\epsfig{file=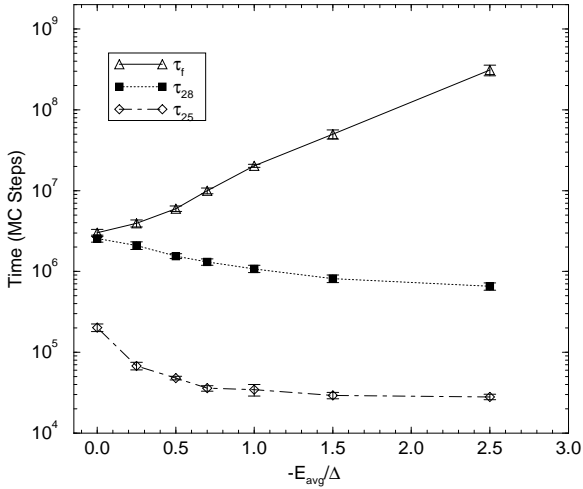,width=\figurewidth}}
\caption{The minimum folding time and the compaction times plotted as a
  function of $E_{\rm avg}/\Delta$ for sequence 002. The line with the
  triangles is the mean folding time (i.e., the time to find the
  native state).  The squares are the mean time to find the first cube
  state (not necessarily the native state) and the diamonds are the
  mean time to find a conformation with 25 contacts. For each of these
  times, there is a minimum point in the point of time vs temperature
  (see fig.~\protect\ref{fig:arren1}). It is that minimum (fastest)
  time that is plotted here for each value of $E_{\rm avg}/\Delta$.}
\label{fig:times002}
\end{figure}

We now explore how the model behaves as we vary the potential
parameters. Specifically, how do the various kinetic and thermodynamic
properties depend on these parameters? Sequence 002 was examined in
detail. As mentioned, this sequence has an unfrustrated cube
conformation (see fig~\ref{fig:cubes}). This cube will be the ground
state whenever $E_l<0$. If $E_l$ is greater than zero then the minimum
energy conformation is the completely unfolded chain with no contacts.
This clearly would not represent a protein under folding conditions;
thus we ignore this region of parameter space. Several simulations
with various values\cite{Note06} of $E_{\rm avg}/\Delta$ ranging from
0 to approximately 2.5 were run.\cite{Note10}
Figure~\ref{fig:times002} shows a plot of the folding and compaction
times versus $E_{\rm avg}/\Delta$.
The times plotted in these figures
are taken from the temperature with the fastest time (i.e., the
minimum point in figure~\ref{fig:arren1}) for each value of $E_{\rm
  avg}/\Delta$.  As the absolute value of $E_{\rm avg}/\Delta$ is
decreased from 2.5 to 0, both compaction times (the time to form 25
and 28 contacts) increase, although the change is small. This is
expected since the drive to form contacts decreases as $\left|E_{\rm
  avg}/\Delta\right|$ is reduced. For the folding time there is an
opposite and more dramatic effect.  Decreasing $\left|E_{\rm
  avg}/\Delta\right|$ decreases the folding time ($\tau_{f}$), almost
two orders of magnitude.  Also, at $E_{\rm avg}=0$, $\tau_{f}$ almost
equals $\tau_{28}$ (the time to make 28 contacts). What is happening
is that as $\left|E_{\rm avg}/\Delta\right|$ decreases, we are
destabilizing non-native cubes relative to the native state. At large
$\left|E_{\rm avg}/\Delta\right|$, these low energy non-native cubes
behave as traps slowing down the folding rate.  Reducing $\left|E_{\rm
  avg}/\Delta\right|$ increases their energy relative to the native
cube eliminating them as traps. This in turn increases the folding
rate.  Alternatively, as $E_{\rm avg}/\Delta\rightarrow0$, $E_u$
increases until it becomes positive. At this point making weak
(incorrect) contacts is unfavored relative to breaking them. This
keeps the chain from forming these weak, incorrect contacts which
would trap it in states different from the native state. The chain is
more effectively funneled into the native conformation; i.e., the
first cube made is almost always the native one. By varying $E_{\rm
  avg}/\Delta$ we can control the time-scale separation between
folding and collapse to maximally compact states.

\begin{figure}[tb]
\centerline{\epsfig{file=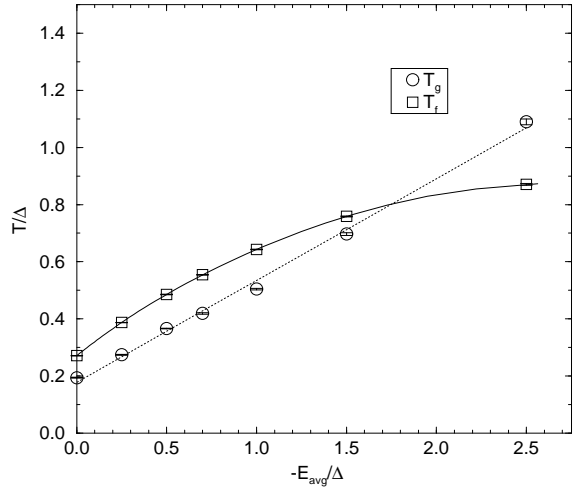,width=\figurewidth}}
\caption{A plot of the folding temperature ($T_f$)
  and the kinetic glass temperature ($T_g$) as a function of the
  average drive to compactness ($E_{\rm avg}$). Both the temperatures
  and the energies have been scaled by $\Delta$.  The dotted line is a
  linear fit of the glass temperatures.  The solid line is not a fit
  to the $T_f$ points but was calculated explicitly via the histogram
  technique. For values of $E_{\rm avg}/\Delta$ approximately less
  than 1.7 $T_f>T_g$ so the chains will fold before hitting the glass
  transition. For $E_{\rm avg}/\Delta$ greater than 1.7 the glass
  transition occurs before folding so the chains do not reach their
  ground state within the simulation time.}
\label{fig:TgTf}
\end{figure}

Next, we examined how the folding temperature ($T_f$) and the kinetic
glass temperature ($T_g$) varied as a function of $E_{\rm
  avg}/\Delta$. In particular, for which values of $E_{\rm
  avg}/\Delta$ is $T_f$ greater than $T_g$?  For these values of
$E_{\rm avg}/\Delta$, the chain will fold before the native state
becomes inaccessible. The histogram method is used to calculate $T_f$
for various values of $E_{\rm avg}/\Delta$. The technique is the same
as the one used to extrapolate to different temperatures. In this case
one needs histograms as a function of good and weak contacts (which
can be calculated from the energy-contact histograms previously used).
To calculate $T_g$ we must run simulations at various values of
$E_{\rm avg}/\Delta$ since there is no way to extrapolate as in the
case of $T_f$. The two temperatures ($T_f$ and $T_g$) are plotted for
sequence 002 in figure~\ref{fig:TgTf}.
Note that we plot the
temperature normalized by $\Delta$ (i.e., in units of $\Delta$) just
as we plot $E_{\rm avg}$ normalized by $\Delta$.\cite{Note10} The
glass temperature varies almost linearly with $E_{\rm avg}/\Delta$.
Previous work on glass transitions in heteropolymers have shown that
the transition occurs after the collapse of the
system.\cite{Bryngelson90,Note11} In fact it was shown that the
polymer needs to collapse in order to have a glass transition. This is
consistent with the behavior we see: as $\left|E_{\rm
  avg}/\Delta\right|$ increase so does the collapse temperature
($T_c$).  In fact as we will see shortly $T_c>T_g$ for all values of
$E_{\rm avg}/\Delta$.  As $\left|E_{\rm avg}/\Delta\right|$ increases,
the depth of the local minima increases relative to the unfolded
state. It becomes harder to escape from local traps so the chain
``freezes'' at higher temperatures.

The folding temperature also increases as the absolute value of
$E_{\rm avg}/\Delta$ increases but reaches a limiting value. This
behavior of $T_f$ with $E_{\rm avg}$ is easy to understand. As
$\left|E_{\rm avg}/\Delta\right|$ is increased, the stability gap
becomes larger; hence $T_f$ increases.  However, it eventually
asymptotes to a limiting value of approximately 0.87.  This limiting
value turns out to be precisely the folding temperature that would be
calculated using just the cube conformations. In
table~\ref{tab:cubevsfull} we see that for sequence 002 $T_f=1.763$
for the cube-only calculation. In calculation $\Delta=2$ so we need to
divide the temperature by 2, giving 0.88.  The reason $T_f$ approaches
the cube-only value is that as $\left|E_{\rm avg}/\Delta\right|$
increases the cube states decrease in energy more than the non-cube
states since they have more contacts.  Eventually, at low enough
$E_{\rm avg}/\Delta$ all the low energy states are just cubes. It is
the low energy states that determine $T_f$. Once $T_f$ reaches this
limit, it is unchanged by further changes to $E_{\rm avg}/\Delta$
since the relative energy between the various cube states is
determined by $\Delta$ which we are holding constant at 1.

\begin{figure}[tb]
\centerline{\epsfig{file=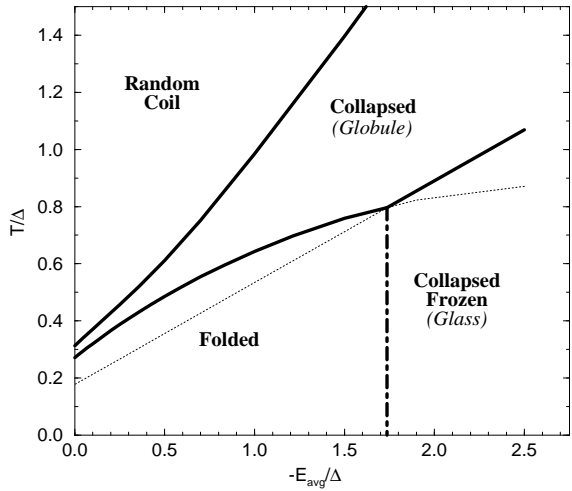,width=\figurewidth}}
\caption{Phase diagram for sequence 002. There are four regions:
  random coil, collapsed globule, collapsed frozen, and folded. The
  solid line between the random coils and the collapsed globule state
  is the collapse transition, the temperature at which half the chains
  have 20 contacts. The second solid line is equal to either $T_f$ or
  $T_g$, whichever is greater at that value of $E_{\rm avg}/\Delta$.
  The dotted line equals the lesser of $T_g$ or $T_f$. The vertical
  dash-dotted line shows the transition from the folding region (where
  $T_f>T_g$) to the frozen region. It occurs at $E_{\rm
    avg}/\Delta\approx-1.7$.  }
\label{fig:phase}
\end{figure}

Looking at both the $T_f$ and $T_g$ lines we see there is a key point
at which $T_g$ becomes greater than $T_f$. At $E_{\rm
  avg}/\Delta\approx 1.7$ the chains become glassy before they become
thermodynamically stable; consequently, they will not be able to fold.
For $E_{\rm avg}/\Delta$ greater than 1.7, $T_g>T_f$. The chain is
trapped in local minimum and will not find the native state within the
simulation time. When $E_{\rm avg}/\Delta$ is less than 1.7 $T_g<T_f$,
so the chains fold to the native state and will be thermodynamically
stable before the dynamics slow down.  By including the results for
the collapse transition temperature (the temperature at which half the
chains have at least 20 contacts) a qualitative phase diagram can be
drawn (see fig.~\ref{fig:phase}).
There are four regions: random
coil, collapsed globule, folded and collapsed frozen state. The phase
diagram is very similar to other lattice models and theoretical
calculations of heteropolymers.\cite{Shakhnovich89,Dinner84} The
vertical dotted line represents the transition from the glassy to
folded phase. To the right of it (i.e. large $\left|E_{\rm
  avg}/\Delta\right|$) $T_f/T_g<1$ and to the left of the line (small
$\left|E_{\rm avg}/\Delta\right|$) $T_f/T_g>1$.  As $\left|E_{\rm
  avg}/\Delta\right|$ is decreased the folding and collapse curves
converge. This is the same behavior observed in the kinetic data (see
fig~\ref{fig:times002}). As $E_{\rm avg}/\Delta$ approaches zero,
$E_u$ becomes positive. Collapsed states with weak contacts will be
unfavored relative to states with no contacts.  This drives the
polymer to form only correct (good) contacts, so the chains will
collapse almost directly to the correctly folded state. Another way to
understand this is that non-native cube conformations are unfavored as
$E_u$ increases.  At $E_{\rm avg}/\Delta=0$ roughly half of the cubes
for sequence 002 will have positive energies. This removes them as
possible kinetic traps. For all values of $E_{\rm avg}/\Delta$, the
collapse temperature is greater than not only the folding, but also
the glass, temperature. Previous heteropolymer mean-field
calculations\cite{Bryngelson90} show that $T_g$ must be less than
$T_\theta$, the collapse temperature, consistent with what we find
here. Perhaps most interestingly, we see that by modulating a single
parameter in our model we obtain a range of qualitatively different
folding behaviors.  Recently, work has been done on the classification
and examination of the various possible folding regimes with a
comparison to experimental data.\cite{Bryngelson95}

\section{CONCLUSIONS}

We have continued our comprehensive analysis of the 27 monomer cube
lattice heteropolymer. The Monte Carlo histogram method proved
extremely useful, allowing us to determine the density of states for
this system and then to calculate a broad range of thermodynamic
quantities. In particular, the method overcame the problem of
dynamical slowing down at low temperatures.  Like many other
heteropolymer studies (both analytical and numerical) we find two
different transitions: a collapse transition with a roughly
sequence-independent collapse temperature and a folding (to the native
state) transition with a sequence-dependent folding temperature. The
collapse transition has a second order-like behavior and the folding
transition seems first order-like. The good folding sequences, i.e.,
the sequences that are stable and have fast folding times, have
sharper, more clearly defined transitions, as viewed from the
temperature dependence of the average energy and the specific heat.
Combining kinetic and thermodynamic data, we studied the unfolding
behavior of the system. The unfolding rate has a much simpler
temperature dependence than the folding rate. The rate varies roughly
linearly with $1/T$ for a broad range of temperature up to the glass
point.

After the density of states was obtained we were not only able to
extrapolate to different temperatures but also to different parameter
values of the energy function. The systems were examined as a function
of the average drive toward compactness, where the average drive is
the average of the two contact energies divided (normalized) by their
difference ($E_{\rm avg}/\Delta$). There is a specific value for this
energy drive at which the kinetic glass temperature became greater
than the folding temperature, indicating that the system would no
longer be able to fold within the simulation time due to trapping. We
constructed a phase diagram as a function of this average drive and
temperature (also normalized by the splitting). As the average energy
drive is reduced, the two transitions, collapse and folding, converge.
At zero-average drive the system collapses almost directly into the
native state.

One criticism of these models is that they are too simple to represent
real proteins. However, even in this simple model we see a broad and
diverse range of behaviors depending on the parameters used. It seems
likely that some of the behaviors of real proteins can be explained by
some particular set of parameters. More importantly, it may well be
the case that different proteins have different folding behaviors.
Some proteins may fold extremely rapidly to the native state,
literally collapsing into the native state, while other proteins may
have a clear separation in time scales between collapse to a compact
but non-native ensemble of structures and the rearrangement of the
chain to the final native form. In our model we can interpolate
between these two regimes by modulating one parameter. In the future,
more realistic models that are still simple enough for a through
analysis may reveal more about the properties and functions of real
proteins.

\acknowledgments

We would like to gratefully acknowledge the computational assistance
of A. Schweitzer, W. Bialek and the NEC Research Institute. We thank
K.~A. Dill, H.~S. Chan and P.~G. Wolynes for interesting and helpful
discussions. We also thank J. Song and A. Schwartz for careful
reading of and enlightening comments on the manuscript.  N.~D.~S. is a
Chancellor's Fellow at UCSD. J.~N.~O. is a Beckman Young Investigator.
This work was funded by the Arnold and Mabel Beckman Foundation and by
the National Science Foundation (Grant No.\ MCB-9316186).


\end{document}